\documentclass[hidelinks,twocolumn, prl, superscriptaddress,a4paper]{revtex4}

\usepackage{lipsum}
\usepackage{braket}
\usepackage{amssymb}
\usepackage{amsfonts}
\usepackage{graphicx}
\usepackage{dcolumn}
\usepackage{orcidlink}
\usepackage{csquotes}
\usepackage{float}
\usepackage{amsmath}
\usepackage{bm}
\begin{document}

\title{State Dependent Spread Complexity Dynamics in Many-Body Localization Transition}

\author{Maitri Ganguli\,\orcidlink{0009-0009-4701-2459}}
\email[]{maitrig@iisc.ac.in}
\affiliation{Department of Physics, Indian Institute of Science, Bangalore}

\author{Aneek Jana\,\orcidlink{0009-0001-1097-4250}}
\email[]{aneekjana@iisc.ac.in}
\affiliation{Center for High Energy Physics, Indian Institute of Science, Bangalore}

\date{\today}

\begin{abstract}
We characterize the Many-Body Localization (MBL) phase transition using the dynamics of spread complexity and inverse participation ratio in the Krylov space starting from different initial states. Our analysis of the disordered Heisenberg spin-1/2 chain unravels that the ergodic-to-MBL transition can be determined from the transition of the pre-saturation peak in the thermofield double state (TFD) spread complexity. On the other hand, if an initially ordered state or a superposition of a small number of such states is chosen, then the saturation value of spread complexity and Krylov inverse participation ratio (KIPR) can distinguish the ergodic phase from the integrable phases, with no sharp difference between the integrable phases. Interestingly, the distinction between the disorder-free integrable and the MBL integrable phase is established by the spread complexity study of random states chosen from unitary and orthogonal Haar ensembles. We also study the complexity dynamics by coupling the system to a bath, which shows distinctive profiles in different phases. A stretched exponential decay of KIPR is observed when the MBL system is connected to the bath, with the decay starting at an earlier time for a greater value of environmental dephasing. Our work sheds light on the efficacy of Krylov space dynamics in understanding phase transitions in quantum many-body systems.
\end{abstract}

\maketitle

\vskip
\baselineskip
\vskip
\baselineskip

\vspace{1.5cm}

\paragraph*{\textbf{Introduction}.---}

In recent years, the study of quantum complexity quantified on the Krylov basis has gained significant interest for its usefulness in understanding the various
aspects of quantum many-body systems, quantum field theories, holographic models, quantum circuits etc \cite{Nandy:2024htc,Hashimoto_2023,Takahashi:2024hex,PhysRevB.106.195125,alishahiha2024thermalizationkrylovbasis,Rabinovici:2023yex,PhysRevD.104.L081702,Avdoshkin_2024,Kundu_2023,suchsland2023krylovcomplexitytrottertransitions,Bhattacharya:2024szw}. The basic notion of complexity captures the difficulty of preparing a certain quantum state starting from a given initial state ~\cite{Nielsen:2006cea}. In the context of quantum dynamics, Krylov Complexity measures the average position of a time-evolved state/operator in the Krylov basis formed by the action of the generator of time-evolution using the Lanczos algorithm or some modified version, such as the bi-Lanczos algorithm in the cases of non-unitary dynamics~\cite{Bhattacharya:2023zqt, Nandy:2024htc}. The Hamiltonian is rendered into a tridiagonal form in the Krylov space spanned by the Krylov basis vectors. Therefore, the complex quantum mechanical state/operator dynamics problem is effectively reduced to an equivalent single particle hopping problem in a semi-infinite lattice numbered by the index of Krylov basis vectors, where the hopping amplitudes are given by the Lanczos coefficients, which are the outputs of the Lanczos/bi-Lanczos algorithm.

While operator complexity in the Krylov space has been studied extensively, in a plethora of both closed and open quantum systems \cite{Rabinovici_2021,Bhattacharjee_2023,Chapman:2024pdw,PhysRevResearch.5.033085,Bhattacharya:2022gbz,Bhattacharjee_2024,Bhattacharjee_2022,Caputa_2024,Bhattacharya:2023xjx}, starting from the seminal paper by Parker et al.\cite{Parker_2019}, the study of the complexity of spread of states, also known as Krylov spread complexity, is relatively new \cite{Balasubramanian:2022tpr,alishahiha23,Bento_2024,Bhattacharya:2023yec,caputa2023krylovcomplexitymodularhamiltonian} and its relevance has been investigated in various contexts including integrability to chaos transitions \cite{Camargo:2024deu, alishahiha2024krylovcomplexityprobechaos} and \(\mathcal P \mathcal T\)-symmetric non-Hermitian Hamiltonians \cite{Bhattacharya:2024hto}, etc very recently. This work aims to contribute to this endeavor by studying state-dependent spread complexity dynamics in the systems that exhibit Many-Body Localization Transition (MBLT) \footnote{Krylov operator complexity in MBL is studied in \cite{Trigueros:2021rwj}}.

The discovery of many-body localization (MBL), where strong disorder and interaction lead to emergent integrability, is a great example of the violation of Eigenstate Thermalization Hypothesis (ETH) \cite{Srednicki_1994,Deutsch_2018,Nandkishore_2015} beyond integrable systems. In the interacting systems, the presence of disorder \cite{Pal_2010} or quasiperiodicity \cite{Iyer_2013} generically can give rise to MBL (which is a generalization of Anderson localization \cite{Guan2019ABI} to interacting systems); however, in thermodynamic limit and greater than one dimension its stability is a subject of active debate \cite{DeRoeck17,Banerjee19}. Recent experimental findings have provided direct evidence of this breakdown of ergodicity in interacting many-body systems. Specifically, these studies have observed such behavior in various systems involving ultracold atomic fermions \cite{schreiberEtAl}, a chain of trapped ions \cite{Smith_2016}, and also in superconducting circuits \cite{XuEtAl}. In these systems, strong disorder caused them to become localized, thereby preventing them from reaching thermal equilibrium as expected in the absence of such disorder.

Though for both MBL systems and disorder-free integrable systems, the existence of an extensive number of conserved quantities is observed, which gives rise to the absence of level repulsion (hence Poisson level spacing distribution in the energy spectrum \cite{Serbyn_2016}), but from the point of view of the entanglement entropy there is a difference. The MBL systems possess eigenstates showing area-law entanglement entropy, whereas generally, with few exceptions, disorder-free integrable systems show volume-law entangled eigenstates. So, from this perspective, identifying the MBL system from thermal eigenstates is easier than identifying the integrable systems. The localization behavior of MBL prevents the system from exploring all possible states, effectively breaking ergodicity—the principle that a system will eventually explore all accessible microstates if given enough time. We are interested in exploring this behavior of MBL systems through the spread complexity and differentiate the MBL emergent integrability from disorder-free integrable systems in this perspective.

 Recent studies have focused on the spread complexity of the Thermo-Field Double (TFD) state, which is a canonical purification of the Gibbs density matrix, to distinguish chaotic and integrable phases \cite{Camargo:2024deu}, as well as to observe the integrable-to-chaotic transition by treating the peak in the spread complexity as an order parameter\cite{Baggioli:2024wbz}. Our study verifies that the peak in the TFD spread complexity indeed shows a transition in the ergodic-to-MBL crossover. 
However, in the present work, our scope is more focused on the ergodic-to-MBL transition in a more elaborate way by studying the initial state dependence of the spread complexity in various physically relevant scenarios. We comment on how the distinguishability among the phases can be seen in the dynamics of initial states chosen from random Haar ensembles, making a clear distinction between the strong-disorder MBL integrable phase and the integrable phase that exists in no or weak disorder in finite-size systems \footnote{we should mention that in the strict thermodynamic limit, even an infinitesimal disorder can break integrability, see \cite{Modak_2014}}. 

Finally, coupling the system to a bath weakly, we have demonstrated that the MBL system shows stretched exponential decay of the Krylov space localization measure (similar to the decay profile of an initial density pattern \cite{Fischer2015DynamicsOA,PhysRevX.7.011034}).

Apart from the Krylov spread complexity, we have also looked at the Krylov Inverse Participation Ratio (KIPR) to understand the dynamics of various states in the Krylov space clearly. The KIPR gives a good dynamic measure of how a state is localized on the Krylov basis. 
All these complexity measures are ultimately dependent upon the wave-function coefficients of the state in the Krylov basis; nevertheless, they allow us to shed light on different aspects of the dynamics in the Krylov space.\\

\paragraph*{\textbf{The Model and the Method}.---}

In this work, we have considered the paradigmatic model of Many-body Localization \cite{_nidari__2008}\cite{Pal_2010}, the spin-1/2 Heisenberg model with random-field disorder,

\begin{equation}
    H = \frac{1}{2}\sum_i \left(\sigma_i^x\sigma_{i+1}^x+\sigma_i^y\sigma_{i+1}^y + \sigma_i^z\sigma_{i+1}^z\right) + \sum_i h_i \sigma_i^z
\end{equation}

here the random-fields \(h_i\) are sampled from a uniform distribution \([-W,W]\). It is known that by increasing the disorder strength, the system goes through an MBL transition around \(W \approx 3.5\) \cite{Abanin_2019}\cite{Buijsman_2019}. This is an instance of a chaotic-to-integrable transition. Another disorder-free integrable phase exists (at \(W=0\)) and is sustained at a very small value of \(W\) for finite size systems \cite{Modak_2014}. We have tried to characterize both of  these integrable and the ergodic phases through the analysis of Krylov complexity, starting from various initial states. We work in the zero magnetization sector (\(\sum_i \sigma^z_i=0\)) for definiteness and use the periodic boundary condition.

The first complexity measure we use is the Krylov Spread Complexity (KSC). Starting from an initial state \(\ket{\psi_i}\), we calculate the orthonormal Krylov basis vectors \(\{\ket{K_n}\}\) (using Lanczos algorithm for Gram-Schmidt orthogonalization on the set \(\{\ket{\psi_i},H\ket{\psi_i},H^2\ket{\psi_i},\dots\}\)) and then expand the time-evolved state in this basis,
\begin{equation}
    \ket{\psi(t)} = e^{-i H t} \ket{\psi_i} = \sum_n \phi_n (t) \ket{K_n}
\end{equation}

The \(\phi_n(t)\)'s can be caculated numerically or using the Lanczos coefficients \(\{a_n\}\) and \(\{b_n\}\) (see Supplemental Material \ref{supp:I}) in solving the following recursive differential equation,
\begin{equation}
    i \frac{d}{dt}\phi_n(t)= a_n \phi_n(t)+b_n \phi_{n-1}(t)+b_{n+1}\phi_{n+1}(t),
\end{equation}
with the boundary condition given by $\phi_n(0)=\delta_{n,0}$.

If we denote the probability of being in the \(n\)-th basis vector by \(p_n\) then,
\begin{equation}
    p_n(t) = | \phi_n (t) |^2,\,\,\, \sum_n p_n(t) = 1
\end{equation}

The Krylov spread complexity is given by the average position in the Krylov basis,

\begin{equation}
    \mathcal C_{\mathcal K} (t) = \sum_n n p_n (t)
\end{equation}

While the above complexity measures average position, we need something that measures the typical number of basis elements needed for describing the time-evolved state by an entropic notion. For this purpose we use the Krylov entropy,
\begin{equation}
    \mathcal S_{\mathcal K}(t) = - \sum_n p_n(t) \log p_n (t)
\end{equation}

and associated Krylov Entropic Complexity (KEC),
\begin{equation}
    \mathcal C_{\mathcal S} (t) = e^{\mathcal S_{\mathcal K} (t)}
\end{equation}

\noindent As a measure of localization in the Krylov space, one can define the Krylov Inverse Participation Ratio (KIPR),
\begin{equation}
\mathcal I_{\mathcal K}(t) = \sum_n p_n^2(t) \leq 1
\end{equation}

\noindent Where larger values of \(\mathcal I_{\mathcal K}\) will imply localization in Krylov space. \(\mathcal I_{\mathcal K}\) also satisfies the lower bound \(\mathcal I_{\mathcal K}\geq 1/d_{\mathcal K} \geq 1/d\), where \(d\) is the dimension of Hilbert space under consideration and \(d_{\mathcal K}\) is the dimension of Krylov space with \(d_{\mathcal K}\leq d\).

 We also consider the effect of dissipation in this system due to its coupling to the environment via a thermal bath or some measurement apparatus. To do so, we have used an effective non-unitary evolution of the state of the system, specifically, the evolution of a single \textit{quantum trajectory}, which corresponds to the \textit{no-click/no-jump limit} in a suitable post-selection procedure.
 
For simplicity, we consider only two jump operators that couple to the system with a coupling strength or dephasing \(\alpha\). The jump operators are chosen to commute with the total magnetization operator to keep the state in the same magnetization sector \cite{Bhattacharya:2023zqt},
\begin{equation}
    \begin{aligned}
        &L_1 = \sigma_0^x\sigma_1^x + \sigma_0^y\sigma_1^y\\
        &L_2 = \sigma_{L-2}^x\sigma_{L-1}^x + \sigma_{L-2}^y\sigma_{L-1}^y
    \end{aligned}
\end{equation}
\noindent The non-Hermitian Hamiltonian is,
\begin{equation}\label{eq: non-Herm H}
     H' =  H - i\alpha \left(L_1^\dagger L_1+L_2^\dagger L_2\right)
\end{equation}

\noindent To elevate the notion of spread complexity for non-Hermitian evolution, one needs to use the bi-Lanczos algorithm, which reduces to the usual Lanczos algorithm in Hermitian limit \cite{Bhattacharya:2023yec}. Here we have two sets of Krylov basis vectors which are bi-orthogonal, \(\{\ket{P_n}\}\) and \(\{\ket{Q_n}\}\), and we can expand the time-evolved state in these two sets as following,
\begin{equation}
    \ket{\psi(t)} = \sum_n \phi^q_n(t) \ket{P_n} = \sum_n \phi^p_n(t) \ket{Q_n}
\end{equation}
For this scenario, one can modify the notion of probability by introducing additional necessary normalization,
\begin{equation}
    p_n (t) = |(\phi^p_n(t))^*\phi^q_n(t)|/\sum_m |(\phi^p_m(t))^*\phi^q_m(t)|
\end{equation}
\noindent Once the probability of being in the \(n\)-th Krylov basis is defined, the definitions of the various complexity measures are kept unchanged.

In the following sections, we emphasize the complexity dynamics for various initial states and the effect of dissipation in the presence of different disorder strengths \(W\). Further, we study the distinctions between the phases that exist for different ranges of \(W\).\\

\paragraph*{\textbf{Thermofield Dynamics}.---} To understand the complexity dynamics, at first, we have chosen the Thermofield Double (TFD) state, which is an entangled state in the product Hilbert space of the two copies of the same system. If the system has a spectrum \(\{E_n\}\), and corresponding eigenvectors \(\ket{\Psi_n}\),  then the TFD state at inverse temperature \(\beta\) is defined by,
\begin{equation}
    \ket{TFD(\beta)} = \frac{1}{\sqrt{Z_{\beta}}}\sum_n e^{-\beta E_n/2} \ket{\Psi_n}_L \otimes \ket{\Psi_n}_R
\end{equation}
where \(Z_{\beta}\) is the thermal partition function and \(L\) and \(R\) denotes left and right copies in the product Hilbert space, respectively. For this state's time evolution, one can consider only the time evolution of the left copy by \(\mathcal H\), and the right copy does not evolve. So effectively, we do the dynamics of the following Gibbs state,
\begin{equation}
    \ket{\psi_\beta}=\frac{1}{\sqrt{Z_{\beta}}}\sum_n e^{-\beta E_n/2} \ket{\Psi_n}
\end{equation}

By choosing \(\beta=0\), we have done the exact time evolution of this state and the corresponding complexity \(\mathcal C_{\mathcal K}\), with different choices of $W$, which gives a characteristic peak in the \(\mathcal C_{\mathcal K}\) evolution for the chaotic regime, and the peak disappears in the MBL regime.

\begin{figure}[ht]
    \centering
    \includegraphics[width=0.85\linewidth]{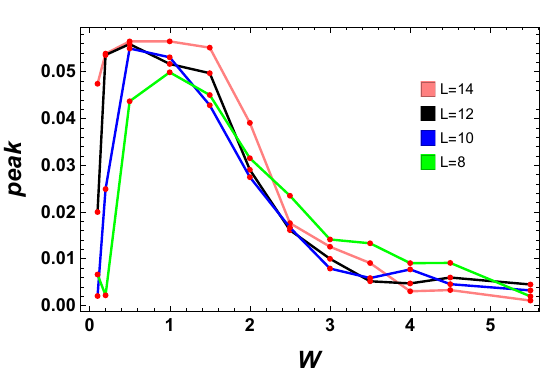}
    \caption{TFD Spread Complexity peak as a function of disorder strength. The plot shows that the transition occurs in the region \(3.5<W<5\), which agrees with the value calculated by level-statistics considerations. The plot also shows that the maximum chaotic feature occurs around \(W\approx 1\), and the transition to integrability starts for \(W>1.5\).}
    \label{fig: peak vs. W TFD}
\end{figure}

The plot of peak height (here we define peak height by (\(\max(\mathcal C_{\mathcal K}(t)/d)-0.5\)), which we take as an order parameter) against disorder strength for various system sizes (see Fig. \ref{fig: peak vs. W TFD}) helps us to determine the range of disorder where the transition occurs, which agrees with earlier estimates from level-statistics transition \cite{Buijsman_2019}.

From Fig. \ref{fig: TFD KC, KIPR, KEC} one important point to note is that the early growth of complexity is controlled by \(W\) monotonically (this can be related to the fact that the minimum energy difference in the spectrum increases with \(W\), \cite{Camargo:2024deu} shows that the time-scale in spread complexity growth depends on the minimum energy difference in the spectrum), even though for larger \(W\), the complexity fails to reach the peak.

 The other two complexity measures shed light on a different aspect of the dynamics in the Krylov space. We observe that (see Fig.\ref{fig: TFD KC, KIPR, KEC}) both the integrable phases, disorder-free integrable and MBL integrable, \textit{delocalizes faster in the Krylov space at early times} than the deep ergodic phase (\(W\approx 1\)). However, the KIPR and the entropic complexity, which measure localization in Krylov space, are the same in the ergodic and MBL phases at late times. In contrast, in the disorder-free integrable phase, the final state is significantly localized in comparison. Understanding the behavior of early time localization in Krylov space of the TFD state can be an important direction for further research.

We have successfully probed the chaotic-to-integrable transition in the context of MBL transitions using the complexity dynamics of the TFD state. However, one should remember that the dynamics of TFD states do not use the information of the associated eigenvectors; it solely depends upon the distribution of eigenvalues. So, studying the dynamics of the TFD state alone will leave us with an incomplete picture of the actual MBL transition. This motivated us to consider the complexity dynamics of other states, which we discuss in the following sections.

\begin{figure}[H]
    \centering
    \includegraphics[width=1.0\linewidth]{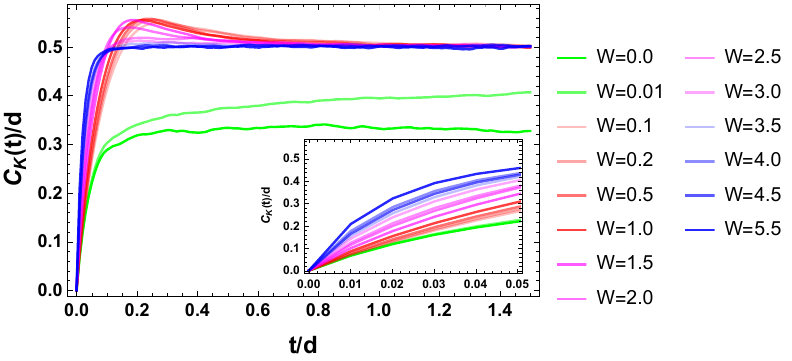}
    \vskip\baselineskip
    \includegraphics[width=1.0\linewidth]{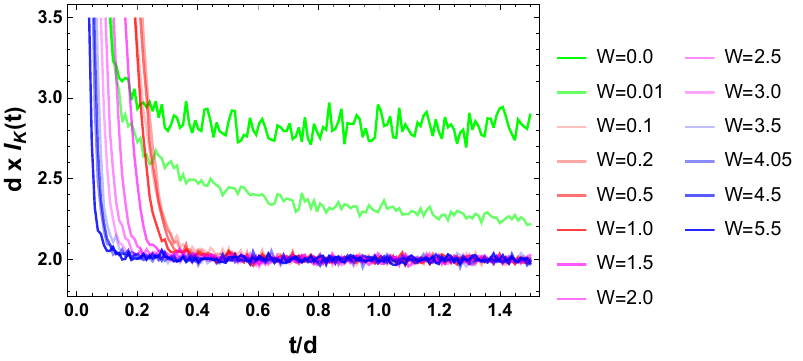}
    \vskip\baselineskip
    \includegraphics[width=1.0\linewidth]{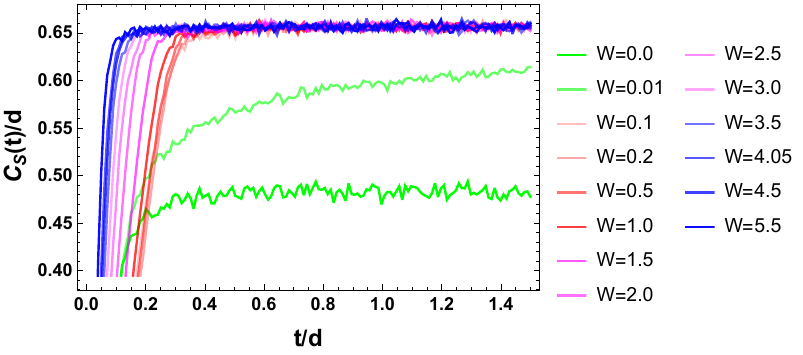}
    \caption{For initial TFD state, \textbf{Top}: Krylov Spread Complexity, \textbf{Middle}: Krylov Inverse Participation Ratio (KIPR), \textbf{Bottom}: Krylov Entropic Complexity. The results are for \(L=14\), where \(d=3432\).}
    \label{fig: TFD KC, KIPR, KEC}
\end{figure}

\paragraph*{\textbf{Relaxation Dynamics}.---}

It is known that MBL systems break ergodicity and can retain some local information about the initial state  \cite{Abanin_2019} \textcolor{blue}. Since our complexity measures essentially give us the information on how distant the time-evolved state is from the initial state and its spread in the Krylov basis, it does make sense to use the spread complexity of states as a probe to measure this memory-retaining property of the MBL phase and its initial state dependence \footnote{In \cite{Cohen:2024ngg}, the authors studied memory-retaining property of MBL phase for different ordered initial states using the statistics of Lanczos coefficients.}.

\begin{figure}
    \centering
    \includegraphics[width=0.95\linewidth]{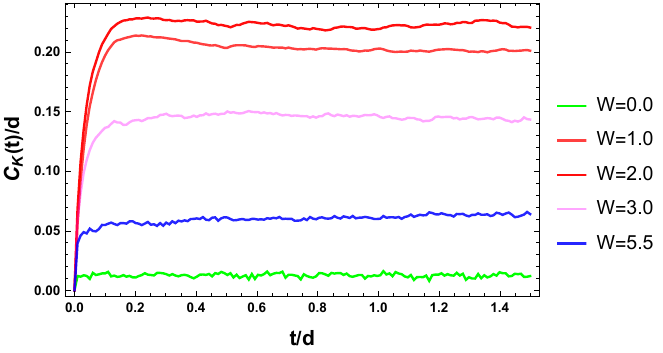}
    \caption{Krylov spread complexity for N\'eel state for different disorder strengths. Result plotted for \(L=12\) with \(d=924\).}
    \label{fig: Neel spread comp. L12}
\end{figure}

For the model under consideration, we prepare the initial state in the computational basis, which has a N\'eel-like order \footnote{The authors in \cite{Gautam:2023bcm} considered the domain-wall-like state as the initial state and found different complexity behavior in ergodic and MBL phases. Our results are more general and explain their observations.},
\begin{equation}\label{eq: neel state}
    \ket{\psi_i} = \ket{1010\cdots10}
\end{equation}
where \(1 (0)\) at \(i\)th position denotes the eigenstate of local \(\sigma_i^z\) operator with eigenvalue \(+1 (-1)\). The results discussed below hold whenever the number of computational basis elements on which the initial state has support, is much less than the total Hilbert space dimension. We show the complexity dynamics for the initial state in Eq.(\ref{eq: neel state}) in Fig. \ref{fig: Neel spread comp. L12}.

Fig. \ref{fig: Neel KC, KIPR, KEC saturation L12} shows that the late-time averaged value of all three complexity measures point towards the fact that the MBL system retains significant memory of the initial state (Robustness of this behavior of retaining the initial memory will be probed by coupling the system to the environment, in a later section). On the other hand, the ergodic phase makes the state more complex and delocalized than its integrable neighbors, as clear from its higher complexity and lesser KIPR.

The above observations can be explained by the existence of pseudo-spin-like quasi-local integrals of motions (LIOMs) \cite{Huse2013PhenomenologyOF} or local-bits (\(\ell\)-bits) \(\tau_i^z\), which have finite overlap with the local spin operators \(\sigma_i^z\) in the presence of strong disorder. Since LIOMs are conserved quantities, information encoded in their initial values remains intact unless the system is coupled to a bath. Therefore, the computational basis elements, eigenstates of the local spin operators \(\sigma_i^z\), show less complex and more localized dynamics in the MBL phase. This is the same reason why, in the MBL phase, the operator Krylov complexity of local \(\sigma_i^z\) operators show more localized behavior (in Krylov space) \cite{Trigueros:2021rwj}.

\begin{figure}
    \centering
    \includegraphics[width=0.8\linewidth]{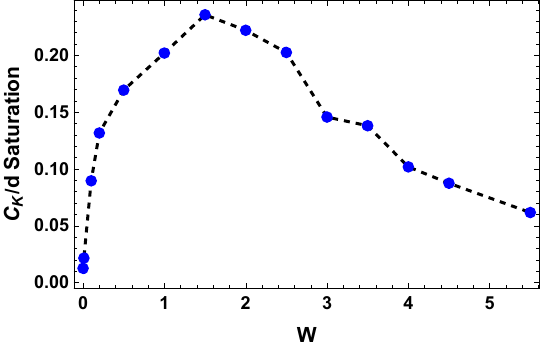}\hfill
    \vskip\baselineskip
    \includegraphics[width=0.8\linewidth]{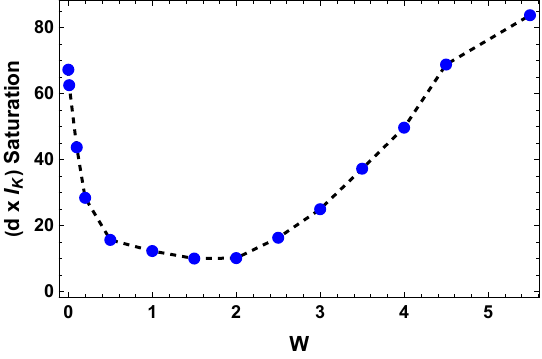}
    \vskip\baselineskip
    \includegraphics[width=0.8\linewidth]{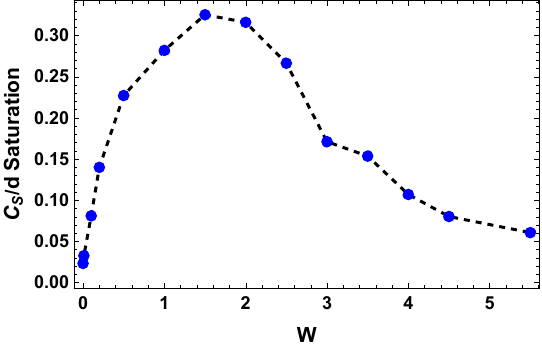}
    \caption{Plots showing how the average saturation value of the various complexity measures of the N\'eel state depend on the disorder strength (we have used \(L=12\), where \(d=924\)). The disorder-free integrable and MBL integrable phases have similar properties for the N\'eel state evolution in terms of higher localization in Krylov space and lesser spread and entropic complexities.}
    \label{fig: Neel KC, KIPR, KEC saturation L12}
\end{figure}

 Now, if we consider extensive superpositions of computational basis elements as our initial state, then the characterization of different phases through complexity becomes more involved. To capture the typical behavior of complexity dynamics in the different phases, we need to be more careful in choosing initial states. To probe typical behavior, we choose Haar random states from two different ensembles, unitary and orthogonal, as our initial states, which we discuss next. \\

\paragraph*{\textbf{Complexity of Typical States}.---} We have been choosing various initial states from the angle of different physical motivations. We have observed that some states, for example, the TFD state and initially ordered states, carry the direct signatures of integrability, be it the disorder-free integrable or MBL integrable phases. Despite these successes, one should try to understand the complexity dynamics of typical states in different phases to see whether the distinctions among the phases via complexity dynamics are generically present.

\begin{figure}[htbp]
    \centering
    \includegraphics[width=0.85\linewidth]{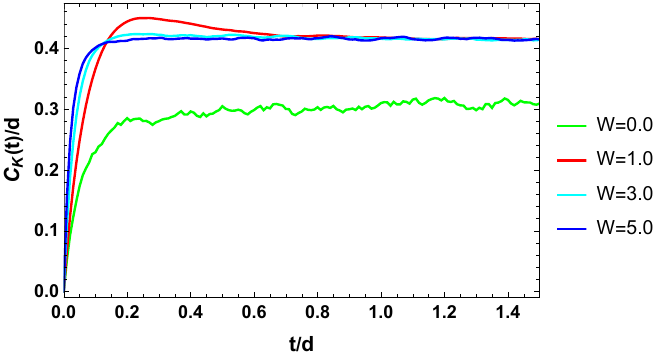}
    \vskip\baselineskip
     \includegraphics[width=0.85\linewidth]{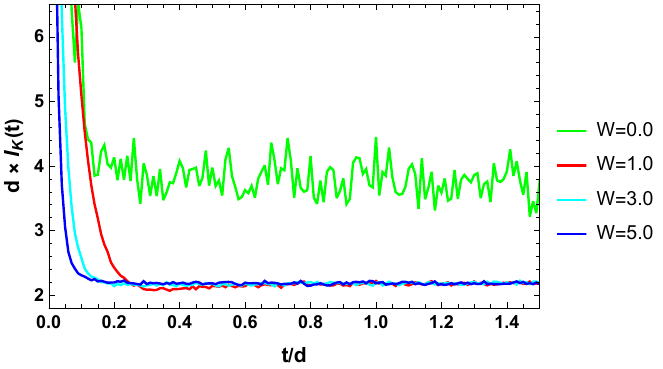}
    \caption{Complexity (top) and KIPR (bottom) of states chosen at random from \(\mathbb C\mathbb P^N\). Spread complexity saturation for \(W\neq 0\) is around 0.4 with a characteristic peak in the ergodic phase profile. The saturation value for the disorder-free phase is around 0.3, and from a KIPR perspective also, this phase is significantly localized. Plots are done for \(L=10\) and \(d=252\).}
    \label{fig: CPN}
\end{figure}

To be completely generic at first, we choose states distributed randomly in the \(N\)-dimensional complex projective space \(\mathbb{C}\mathbb{P}^N\) according to the Haar measure (where (\(N+1\)) is the dimension of the Hilbert space). Such states can be sampled by the action of random \((N+1)\times (N+1)\) unitary matrices on some chosen arbitrary state. We then compute the complexity dynamics of such random Haar states evolving under specified Hamiltonian. 
We find that the MBL phase can be distinguished from the ergodic phase from the absence of a peak in the complexity profile and absence of a dip in the KIPR profile. On the other hand, the disorder-free integrable phase can be distinguished from its lower complexity and higher KIPR, see Fig. \ref{fig: CPN}.
\\

\begin{figure}[htbp]
    \centering
    \includegraphics[width=0.85\linewidth]{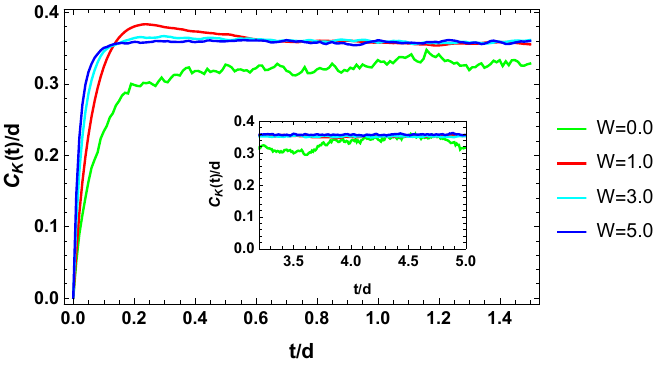}
    \vskip\baselineskip
     \includegraphics[width=0.85\linewidth]{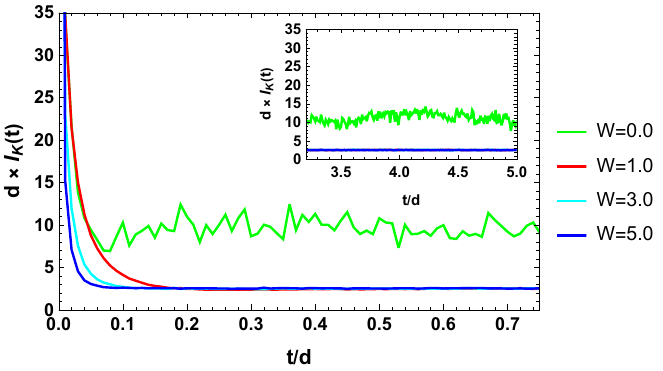}
    \caption{Complexity (top) and KIPR (bottom) of states chosen at random from \(\mathbb R\mathbb P^N\). Inset: Late-time behavior. The late-time complexity value of all three phases is almost the same, but in the Krylov space, the disorder-free phase is more localized. Plots are done for \(L=10\) and \(d=252\)}
    \label{fig: RPN}
\end{figure}

\noindent Another choice can be to sample random states uniformly from the real projective space \(\mathbb R \mathbb P^N\), which can be physically important while being completely random. Such states can be sampled by acting random orthogonal matrices on some arbitrarily chosen state. It is again observed that a peak in the complexity can distinguish the ergodic phase from the integrable ones. Even though, at late times, all of them have similar saturation values, unlike the \(\mathbb C \mathbb P^N\). But from Fig. \ref{fig: RPN}, it is clear that the KIPR value still can set apart the disorder-free integrable phase from the MBL phase.

From these observations, we infer that the distinctions among the chaotic and the integrable phases are not special to states like TFD; rather, they occur in more generic states also. Therefore, it is worth understanding the physical origin of such distinctive complexity behavior of different kinds of random states under the evolution of different Hamiltonians.

\paragraph*{\textbf{Dissipative Dynamics}.---}
MBL is recognized as a robust dynamical phase of matter, when strictly decoupled from the environment, does not thermalize \cite{Abanin_2019}. However, coupling to the environment necessarily leads to a delocalization transition, with a time scale governed by the coupling parameter value. The goal is to understand this transition from a Krylov complexity perspective and point out the robustness of the MBL integrability.

  We model the coupling to a bath or a measurement apparatus by an effective non-unitary evolution by the Hamiltonian in Eq.(\ref{eq: non-Herm H}), which describes a specific quantum trajectory. Under the assumption of weak dephasing (\(\alpha<<1\)), this effective description is justified, and as we will demonstrate, it well captures the essential physics even within this minimal setup. To be specific, we start with the N\'eel state, whose complexity dynamics in a closed system can distinguish the ergodic from the MBL, to see how its evolution is affected by the environmental coupling.
  
\begin{figure}[htbp]
      \centering
      \includegraphics[width=1.0\linewidth]{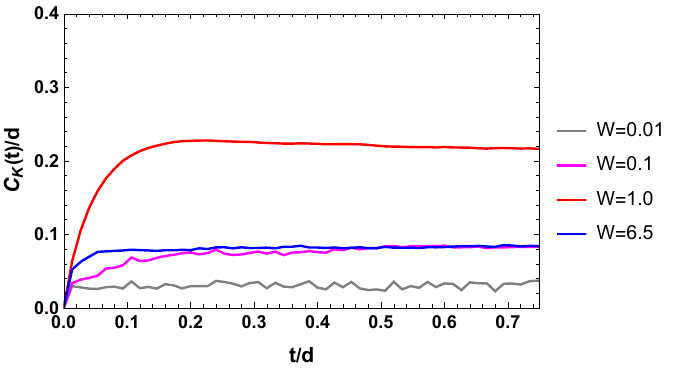}
      \vskip\baselineskip
      \includegraphics[width=1.0\linewidth]{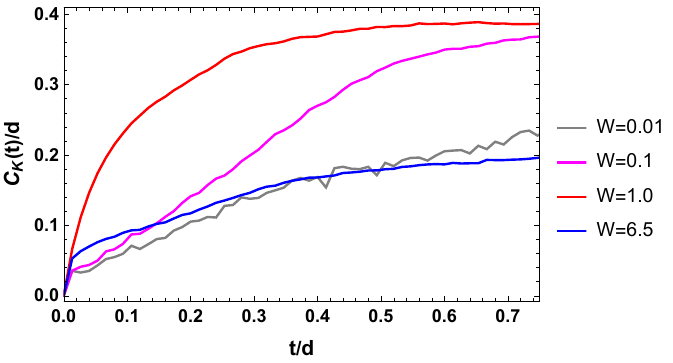}
      \vskip\baselineskip
      \includegraphics[width=0.45\linewidth]{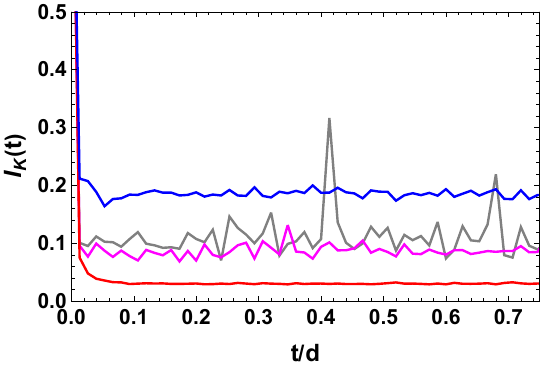}
      \includegraphics[width=0.45\linewidth]{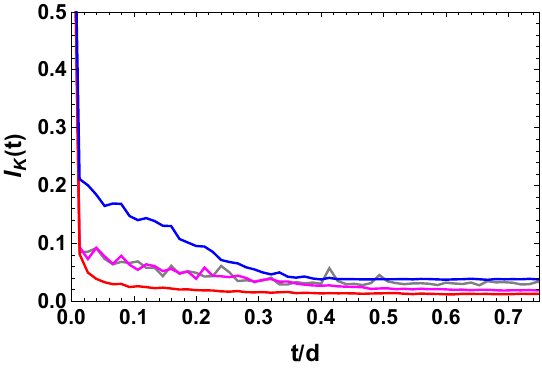}
      \caption{Comparing early-time spread complexity and KIPR dynamics in closed and open systems for the initial N\'eel state. The plots are done for \(L=10\) and \(200\) realizations. \textbf{Top}: spread complexity dynamics for no-environmental coupling, \(\alpha=0\). \textbf{Middle}: spread complexity dynamics for non-zero environmental coupling, \(\alpha=5\times 10^{-3}\). \textbf{Bottom-left}: KIPR for closed system (\(\alpha=0\)), \textbf{Bottom-right}: KIPR for open system (\(\alpha=5\times 10^{-3}\)).}
      \label{fig: open-different W-alpha-0.005}
  \end{figure}

   Spread complexity and KIPR for different disorder strengths \(W\) are affected by non-zero \(\alpha\), which is shown in Fig. \ref{fig: open-different W-alpha-0.005}. It is found that coupling to the environment certainly causes \textit{delocalization in the Krylov space} and an \textit{increase of spread complexity}. We have observed that, by increasing disorder \(W\), the early-time growth rate of spread complexity increases first and decreases by further increasing of \(W\). In particular, in the large \(W\) phase, the early-time growth rate is comparably less \footnote{However, in the late-time spread complexity behavior, we have found that in the large \(W\) phase, the complexity is more than the chaotic and disorder-free integrable phase. To understand the precise reason, further investigation is needed. We would like to address this point in a future work.}. Therefore, the MBL phase is less susceptible to environmental coupling from the Krylov complexity perspective, at least at early times.
  
\begin{figure}[htbp]
      \centering
      \includegraphics[width=0.85\linewidth]{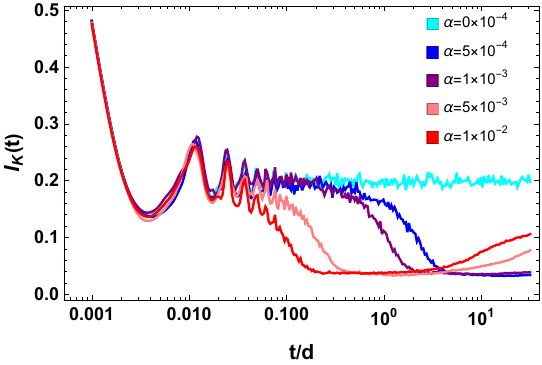}
      \caption{Decay of Krylov space IPR of initial N\'eel state in the MBL phase (with disorder strength \(W=6.5\)) in the presence of environmental coupling with different strengths. For weaker coupling, a steady IPR is maintained for a longer period of time before an exponential delocalization. The plot is done for \(L=10\) with \(250\) disorder realizations.}
      \label{fig: open-diff.alpha-W6pt5}
  \end{figure}

\begin{figure}[htbp]
      \centering
      \includegraphics[width=0.85\linewidth]{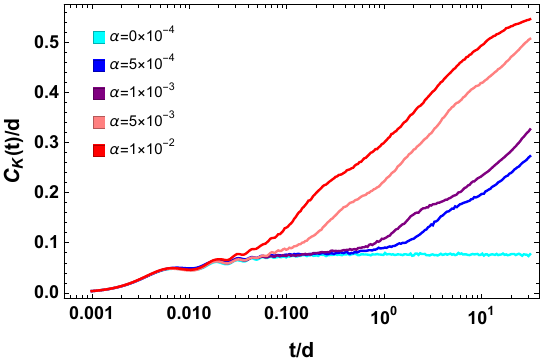}
      \caption{Increment profile of early-time spread complexity of initial N\'eel state when the MBL system is coupled to a bath with different coupling strengths. At very late times, for non-zero \(\alpha\), we have observed that the complexity value saturates above \(0.50\). The plot is done for \(L=10\) with \(250\) disorder realizations.}
      \label{fig: open-diff.alpha-W6pt5-KC}
  \end{figure}

  Now, focusing on the delocalization transition in MBL systems for different values of the environmental coupling \(\alpha\), we have observed that the Krylov IPR has a stretched exponential decay profile (see Fig. \ref{fig: open-diff.alpha-W6pt5}), which is similar to the decay profile of initially set particle density imbalance found earlier in \cite{Fischer2015DynamicsOA} and decay of third Renyi negativity as found in \cite{Wybo_2020}\footnote{In this context, one should note that the idea of KIPR is more robust in the angle of understanding the localization property for the system. We do not need to use N\'eel like order to observe this kind of characteristic decay of KIPR in open quantum systems, in fact, any computational basis state would capture this feature.}.
  However, the complexity increases always for non-zero \(\alpha\) (see Fig. \ref{fig: open-diff.alpha-W6pt5-KC}) with a possible saturation above \(0.5\) at very late times. This indicates that the time-evolved state really goes far from the initial state in the Krylov space at late times. These observations show that the quasi-local integrals of motion (LIOMs) are destroyed when the MBL system is coupled to a bath, which is efficiently captured by Krylov space dynamics.\\

\paragraph*{\textbf{Discussion and Future Directions}.---}
In our work, we have studied the state-dependent spread complexity dynamics in the disordered spin-1/2 Heisenberg chain to understand whether complexity dynamics in the Krylov space can make distinctions among the disorder-free integrable, ergodic, and emergent integrable MBL phases. 

\begin{itemize}
    \item Through our analysis we show that the pre-saturation peak height in TFD state complexity, which we use as an order parameter, can significantly capture the ergodic to MBL transition.
    \item If we start with an initially ordered state like N\'eel state we can also infer different phases from the saturation values of Krylov Complexity and Krylov Inverse Participation Ratio (KIPR).
    \item Our work has established that not only special states like TFD or initially ordered states but also randomly chosen typical states carry important information about the phases in their complexity dynamics.
    \item However for the open system if we focus on the early-time growth of the complexity of the initial N\'eel state, in the MBL phase it is slower than the other two phases. Interestingly, for the MBL phase coupled to a bath, we observe the dissipative delocalization in the Krylov space with a stretched exponenetial decay profile.
\end{itemize}
   
So, our analysis highlights the effectiveness of Krylov space methods in understanding the different phases for both closed and open quantum dynamics.

One important future direction is to do the complexity dynamics in the quasiperiodic systems that show MBL transition, such as the interacting Aubry-Andre model \cite{Iyer_2013,Huang_2024}, to understand if there is any qualitative difference in the complexity dynamics of disordered systems and the quasiperiodic potential systems. Investigating the role of interaction in the complexity dynamics is also interesting. In the spin model, this amounts to tuning a parameter that multiplies the term \(\sigma_i^z\sigma_{i+1}^z\). One should also try to understand if time-reversal symmetry (TRS) has any role in this context. For that, a term that breaks time-reversal symmetry \cite{Buijsman_2019} is to be added to the Hamiltonian. Then, the complexity dynamics should be studied for this Hamiltonian that breaks TRS.

Finally, we leave it for future work to analyze the spread complexity dynamics in the MBL phenomenological model in terms of the LIOMs \cite{Huse2013PhenomenologyOF}. That can potentially provide a more model-independent way of characterizing MBL transition through the spread complexity dynamics.\\

\begin{acknowledgments}

M.G. would like to thank Sumilan Banerjee and Subroto Mukerjee for useful discussions. The authors also thank Aranya Bhattacharya for comments on the draft and suggestions. M.G. is supported by the Integrated PhD fellowship of Indian Institute of Science, Bengaluru, and A.J. is supported by INSPIRE fellowship by DST, Govt. of India.\\

\end{acknowledgments}

\paragraph{Note added.--} After the completion of this work, we became aware that the authors \cite{aranya2024} are investigating complexity in random unitary circuits, which also have signatures of MBL like behavior from a Krylov complexity perspective.

\bibliographystyle{apsrev4-1}
\bibliography{apssamp}

\newpage

\onecolumngrid

\section*{Supplemental Material}

\section{Bi-Lanczos algorithm and complexity}\label{supp:I}

The bi-Lanczos algorithm is suitable for tri-diagonalizing a non-Hermitian matrix (\(H^\dagger \neq H\)). This can be applied for non-Hermitian state evolutions (in non-unitary dynamics, no-click limit in MIPT or no-jump limit in open quantum system evolution) or operator evolution (in the context of open quantum system dynamics and general measurement settings). In the context of state evolution, the Hamiltonian \(H\) will be non-Hermitian, and in the context of operator evolution, the Lindbladian \(\mathcal L_o\) (\(\mathcal L_o^\dagger \neq \mathcal L_o\)) will be non-Hermitian. Once the inner product is specified, the following bi-Lanczos algorithm can be used for both non-Hermitian state and non-Hermitian operator evolution. For Hermitian generators, this reduces to the original Lanczos algorithm. This algorithm is based on \cite{Bhattacharya:2023zqt,Bhattacharya:2023yec}. \\

\noindent Suppose we trying to tri-diagonalize an operator \(M\) which is non-Hermitian, that is \(M^\dagger \neq M\). In this case, we have to construct two sets of Krylov basis vectors \(\{|P_n\rangle\}\) and \(\{|Q_n\rangle\}\), which are bi-orthogonal, 
\begin{equation}
    \langle Q_m | P_n \rangle = \delta_{mn}
\end{equation}

So these two sets are orthogonal with respect to each other, but inside each set, the vectors are not orthogonal; by that we mean \(\langle Q_m | Q_n \rangle \neq \delta_{mn}\) and \(\langle P_m | P_n \rangle \neq \delta_{mn}\).\\

As we will see, we shall generate three sets of Lanczos coefficients, the main diagonal \(\{a_n\}_{n\geq0}\), the upper diagonal \(\{b_n\}_{n\geq1}\) and the lower diagonal \(\{c_n\}_{n\geq1}\). We shall have \(b_n = c_n\) for the Hermitian limit, but they would be different for non-Hermitian.

\paragraph{Constructing the Krylov basis vectors.} Start with \(\ket{P_0} = \ket{Q_0} = \ket{\psi(0)} \text{ or } \ket{\mathcal O_0}\) for state and operator complexity respectively. And define the initial Lanczos coefficients values \(a_0 = \bra{Q_0} M \ket{P_0},\, b_0 = 0,\, c_0 = 0\). The following algorithm is similar to the usual Lanczos algorithm where the \(\ket{P_n}\)'s are constructed using \(M\) and \(\ket{Q_n}\)'s are constructed using \(M^\dagger\) but in a correlated way.\\

\noindent For \fbox{\(n=0\)}, 
\begin{equation}
\begin{aligned}
    &\ket{A_1} = M \ket{P_0} - a_0 \ket{P_0} \\
    &\ket{B_1} = M^\dagger \ket{Q_0} - a_0^* \ket{Q_0}\\
    & w_1 = \bra{A_1}B_1\rangle,\,\, c_1 = \sqrt{|w_1|},\,\, b_1 = \frac{w_1^*}{c_1}\\
    & \ket{P_1} = \ket{A_1}/c_1,\,\, \ket{Q_1} = \ket{B_1}/b_1^*\\
    & a_1 = \bra{Q_1} M \ket{P_1}
\end{aligned}
\end{equation}

\noindent For \fbox{\(n\geq 1\)}, 
\begin{equation}
\begin{aligned}
    &\ket{A_{n+1}} = M \ket{P_n} - a_n \ket{P_n} - b_n \ket{P_{n-1}} \\
    &\ket{B_{n+1}} = M^\dagger \ket{Q_n} - a_n^* \ket{Q_n} - c_n^* \ket{Q_{n-1}}\\
    & w_{n+1} = \bra{A_{n+1}}B_{n+1}\rangle,\,\, c_{n+1} = \sqrt{|w_{n+1}|},\,\, b_{n+1} = \frac{w_{n+1}^*}{c_{n+1}}\\
    & \ket{P_{n+1}} = \ket{A_{n+1}}/c_{n+1},\,\, \ket{Q_{n+1}} = \ket{B_{n+1}}/b_{n+1}^*\\
    & a_{n+1} = \bra{Q_{n+1}} M \ket{P_{n+1}}
\end{aligned}
\end{equation}

It can be shown that the two sets of Krylov basis vectors as formed by above algorithm are indeed bi-orthogonal. In practical purposes one has to stop once \(c_{n+1}\) is less than some cut-off and one has to full-orthogonalize \(|A_{n+1}\rangle\) in the \(\ket{Q_m}\) basis and full-orthogonalize \(\ket{B_{n+1}}\) in the \(\ket{P_m}\) basis for \(m\leq n\).\\

\noindent Observe that,

\begin{equation}
    \begin{aligned}
        M \ket{P_n} &= a_n \ket{P_n} + b_n \ket{P_{n-1}} + c_{n+1} \ket{P_{n+1}}\\
        M^\dagger \ket{Q_n} &= a_n^* \ket{Q_n} + c_n^* \ket{Q_{n-1}} + b_{n+1}^* \ket{Q_{n+1}}
    \end{aligned}
\end{equation}

It shows that \(M\) is rendered to a tridiagonal form in the \(\ket{P_n}\bra{Q_m}\) basis.

\paragraph{Complexities.}

\noindent Now expand \(\ket{\psi(t)} \text{ or } \ket{\mathcal O (t)}\) in the both sets of Krylov basis vectors (let's call them \(P\)-basis and \(Q\)-basis vectors).

\begin{equation}
    \ket{\psi(t)} = \sum_n \phi_n^q (t) \ket{P_n} = \sum_n \phi_n^p (t) \ket{Q_n}
\end{equation}

\noindent where,

\begin{equation}
    \phi_n^q (t) = \bra{Q_n} \psi(t)\rangle, \,\,\phi_n^p (t) = \bra{P_n} \psi(t)\rangle
\end{equation}

\noindent For non-Hermitian evolution we define a probability \(P(t)\) by,

\begin{equation}
    P(t) = \sum_n |(\phi_n^p (t))^* \phi_n^q (t)| 
\end{equation}

\noindent Let's define the Krylov complexity as average position in the Krylov basis dictated by the wavefunction in this basis,

\begin{equation}
   \mathcal C_{\mathcal K}(t) = \frac{\sum_n n |(\phi_n^p (t))^* \phi_n^q (t)|}{\sum_m |(\phi_m^p (t))^* \phi_m^q (t)|} = \sum_n n p_n
\end{equation}

where we have the diagonal probabilities,
\begin{equation}
    p_n = |(\phi_n^p (t))^* \phi_n^q (t)|/\sum_m |(\phi_m^p (t))^* \phi_m^q (t)|
\end{equation}

\noindent The Krylov entropy can be defined as the following Shannon entropy of the diagonal probabilities,

\begin{equation}
    \mathcal S_{\mathcal K}(t) = - \sum_n p_n \log p_n
\end{equation}

\noindent Then we define entropic complexity as exponential of the Krylov entropy,

\begin{equation}
    \mathcal C_{\mathcal S}(t) = e^{S(t)}
\end{equation}

\noindent We also define the Krylov Inverse Participation Ratio (KIPR),

\begin{equation}
    \mathcal I_{\mathcal K}(t) = \sum_n p_n^2
\end{equation}

The KIPR signifies the extent of localization of the wave function in the Krylov basis, and the Krylov entropic complexity signifies the effective number of Krylov basis vectors on which the wave function has typical support.

\section{Level statistics and statistics of Lanczos coefficients}
\begin{figure}[H]
    \centering
    \includegraphics[width=0.3\linewidth]{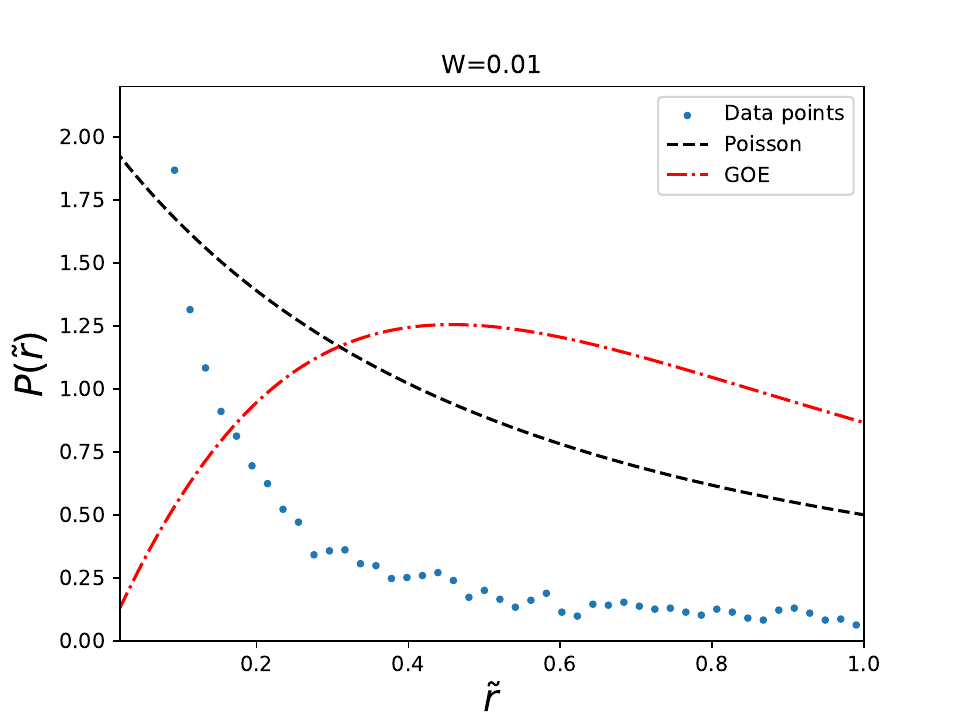}
     \includegraphics[width=0.3\linewidth]{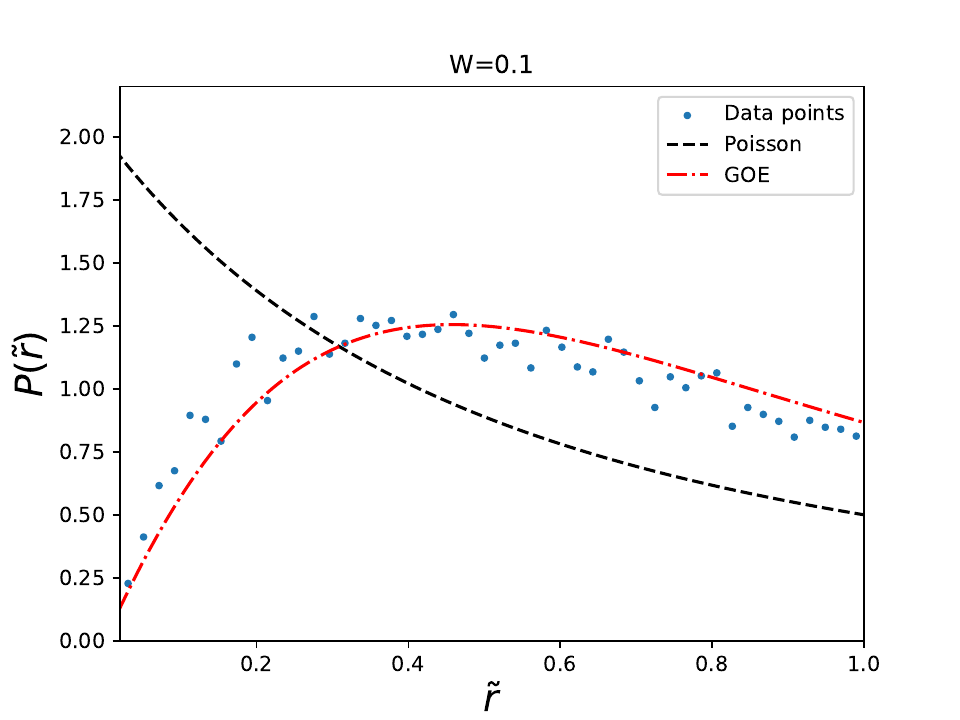}
      \includegraphics[width=0.3\linewidth]{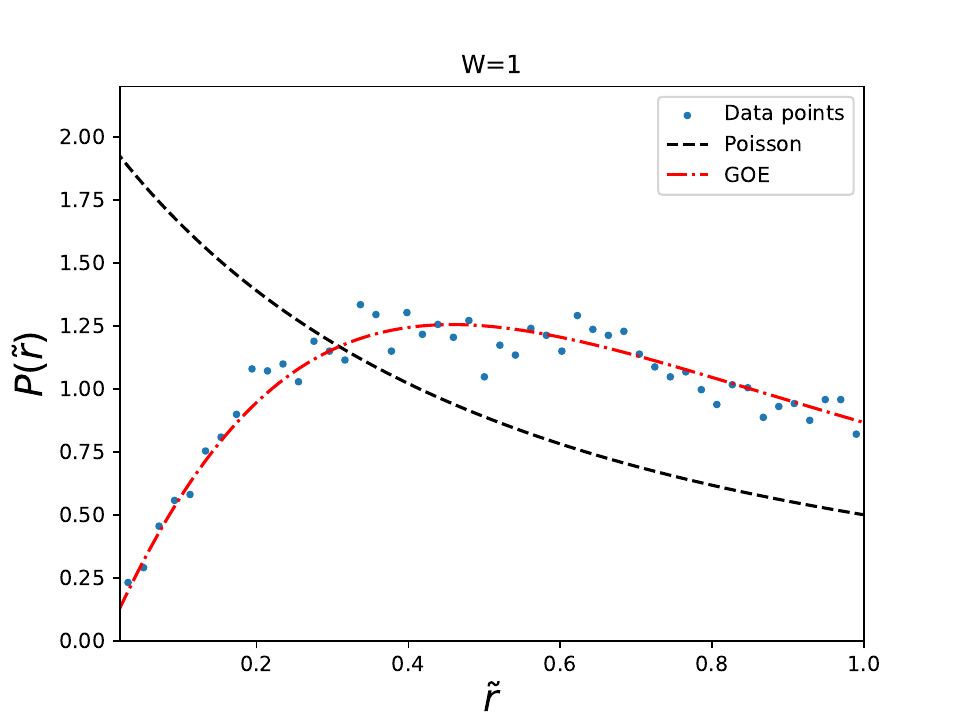}
      \vskip\baselineskip
      \includegraphics[width=0.3\linewidth]{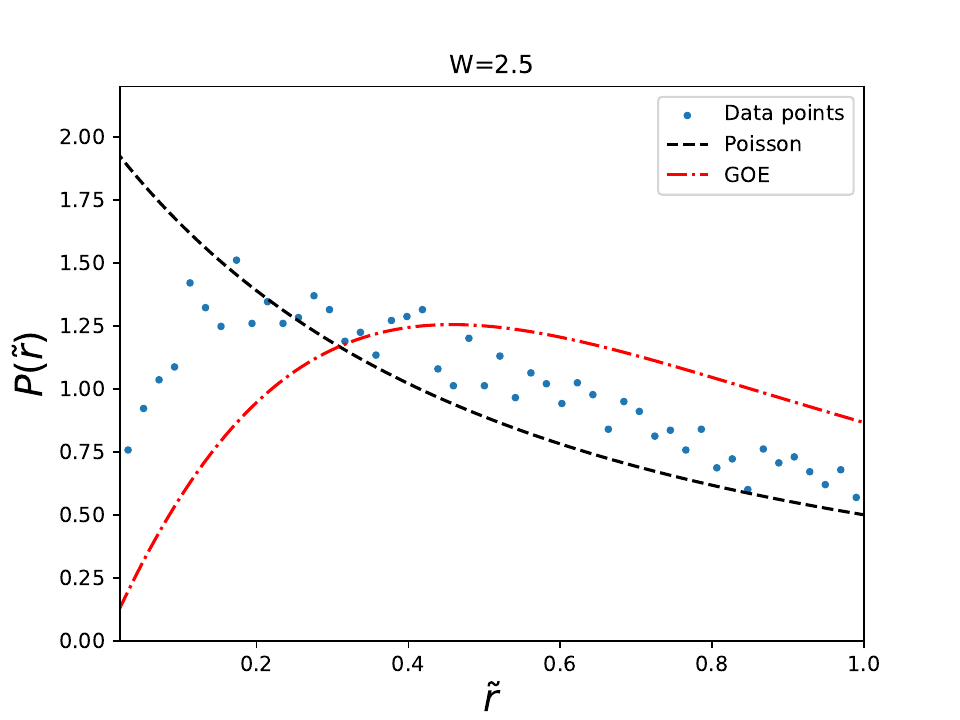}
     \includegraphics[width=0.3\linewidth]{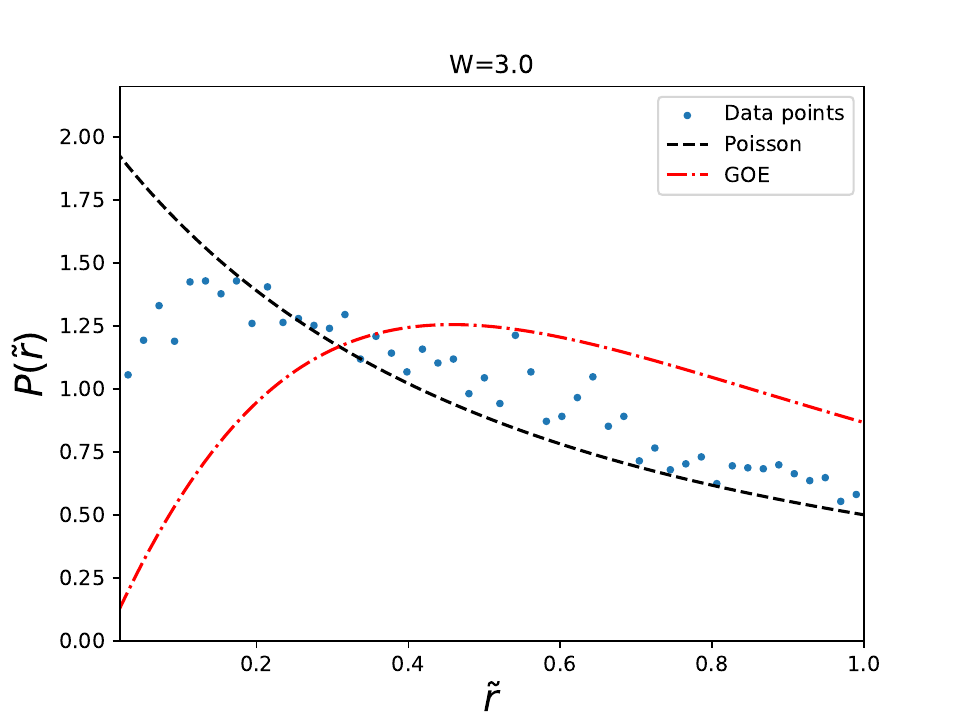}
      \includegraphics[width=0.3\linewidth]{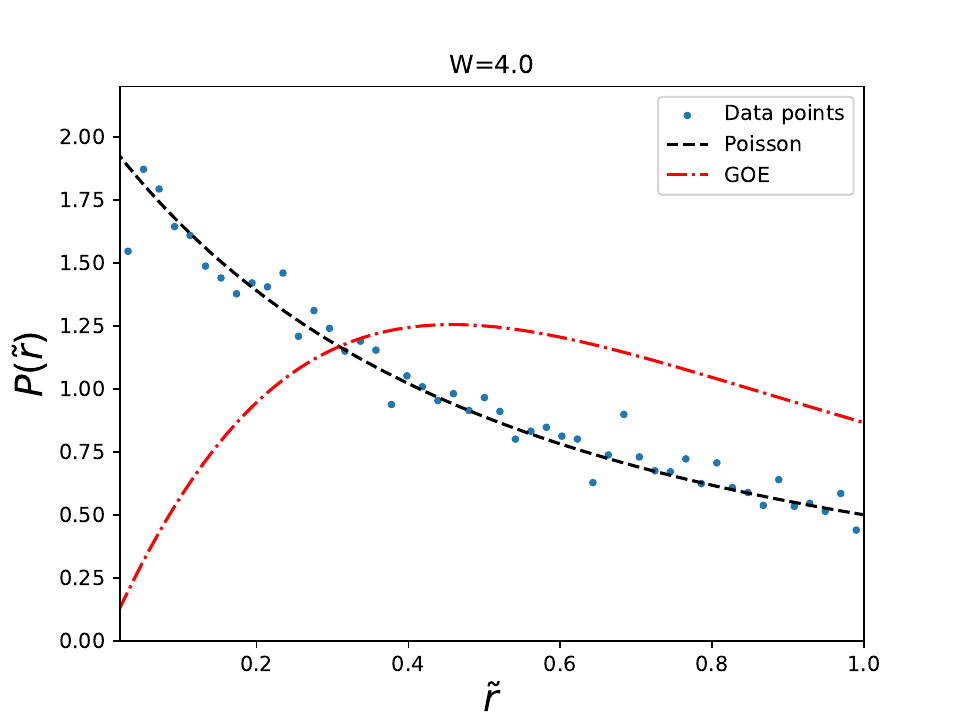}
    \caption{Transition in level statistics by changing disorder strength \(W\) (here, restricted ratio \(\Tilde{r}_n = \min(r_n,1/r_n)\), where \(r_n=(E_{n+1}-E_n)/(E_n - E_{n-1})\) is the level-spacing ratio). The above plots suggest that the ergodic-to-MBL transition occurs in the range \(3.0<W<4.0\). (The data is collected for \(L=14\)).}
    \label{fig: Level stat}
\end{figure}

The transition of level statistics is a useful probe for chaotic-to-integrable transition. For chaotic or ergodic systems, we have the Wigner-Dyson level statistics, and for integrable systems, we have the Poisson level statistics. However, during chaotic to integrable transition, the level statistics change smoothly from Wigner-Dyson to Semi-Poissonian to completely Poisson. A flow can model this whole transition in the space of different random matrix ensembles, specifically Gaussian \(\beta\)-ensembles, by changing the value of \(\beta\) (note that \(\beta=1\) corresponding to Gaussian Orthogonal Ensemble (GOE) and \(\beta=0\) correspond to Poisson distribution) \cite{Buijsman_2019}, see Fig. \ref{fig: Level stat}.

\begin{figure}[H]
    \centering
    \includegraphics[width=0.3\linewidth]{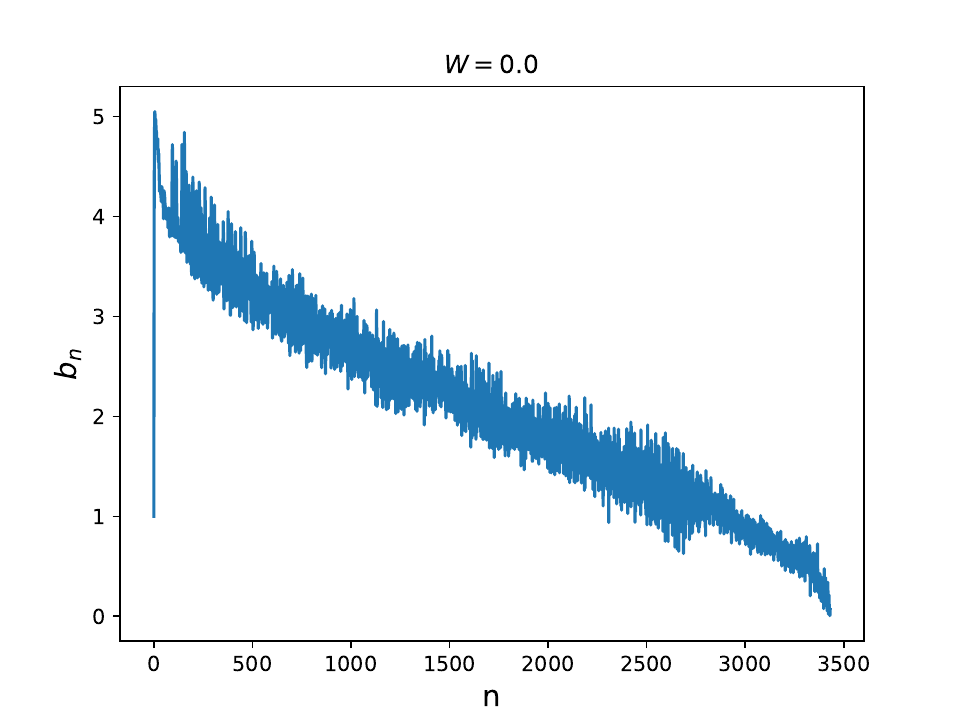}
     \includegraphics[width=0.3\linewidth]{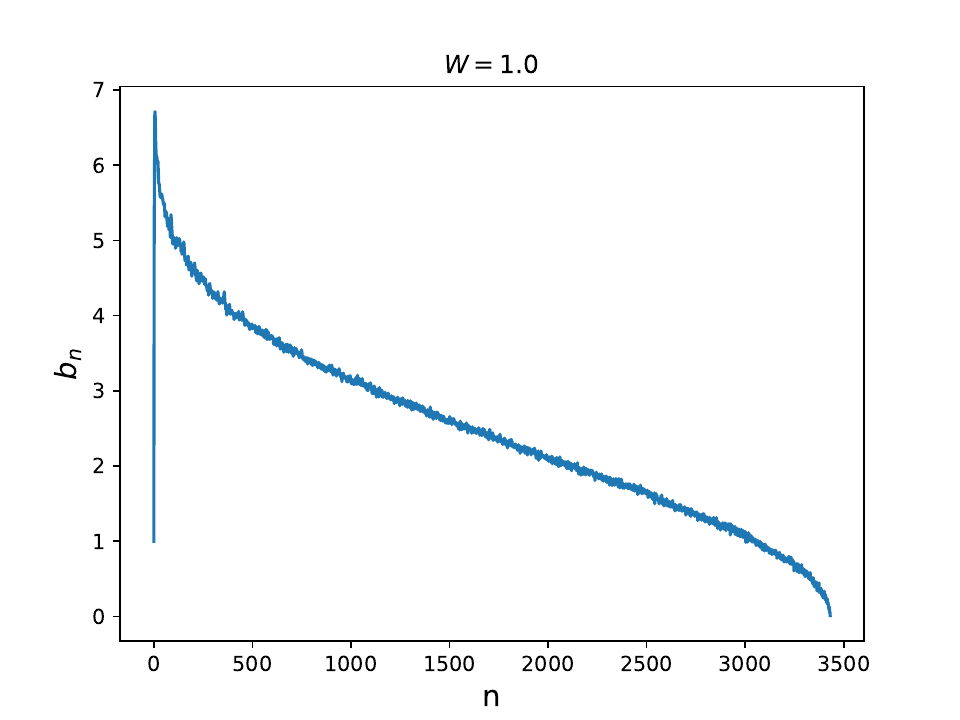}
      \includegraphics[width=0.3\linewidth]{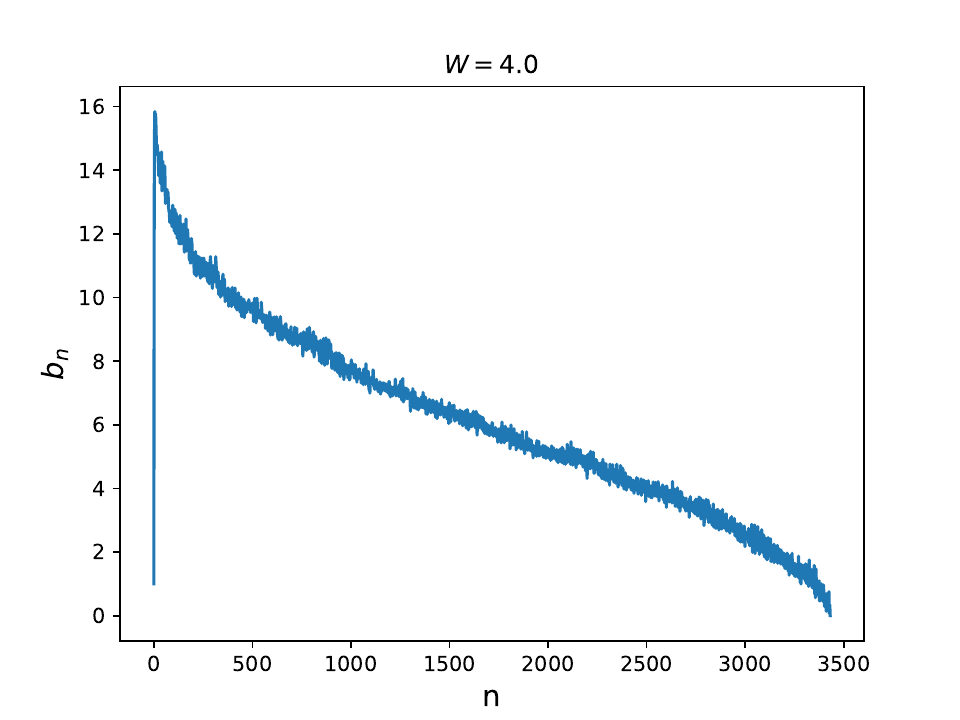}
      \vskip\baselineskip
      \includegraphics[width=0.3\linewidth]{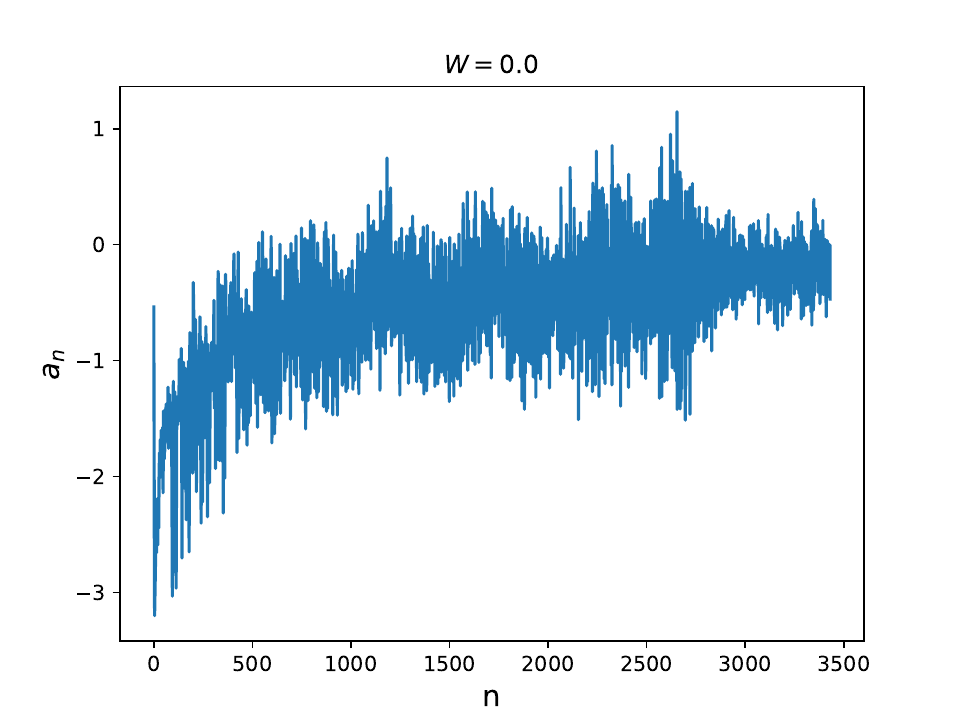}
     \includegraphics[width=0.3\linewidth]{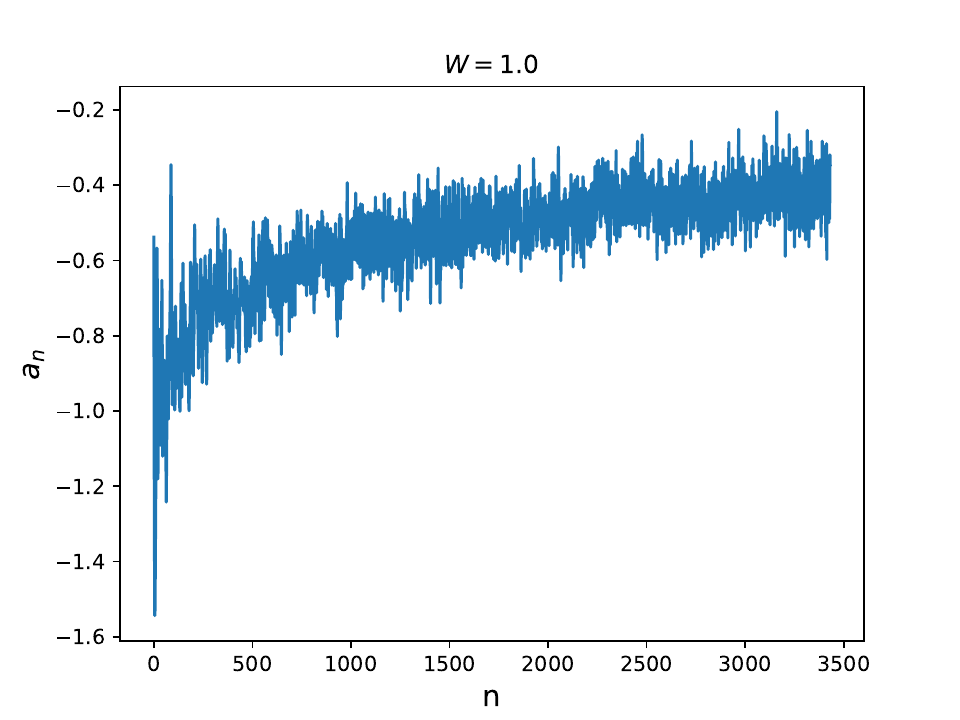}
      \includegraphics[width=0.3\linewidth]{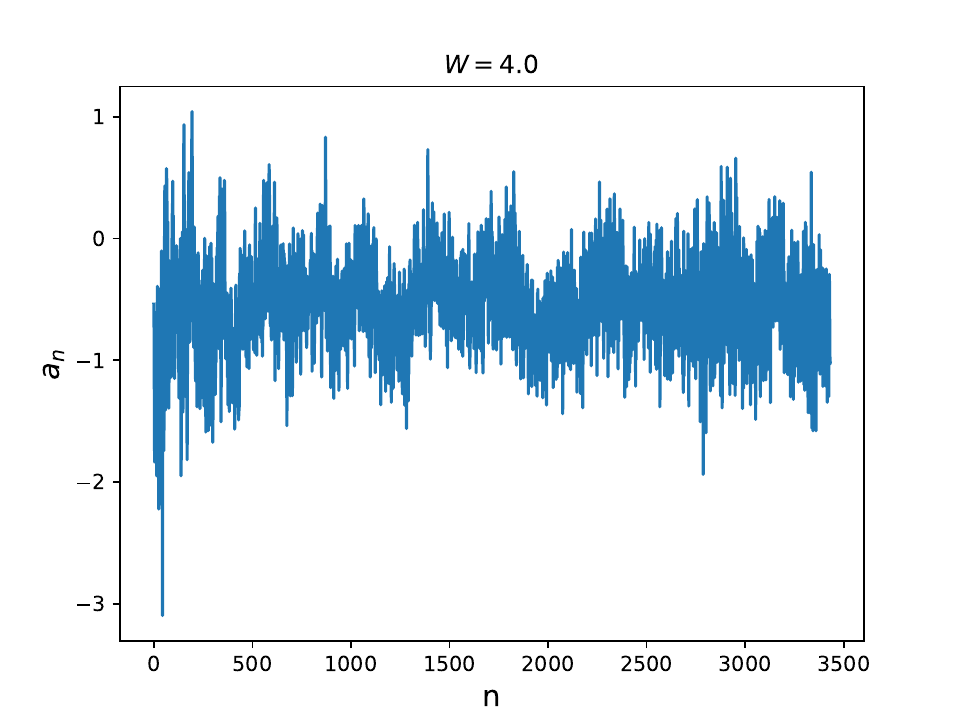}
    \caption{Lanczos coefficients with initial TFD state. Top-panel: \(b_n\) vs. \(n\) for different disorder strengths. Bottom-panel: \(a_n\) vs. \(n\) for different disorder strengths. \(L=14\) }
    \label{fig:Lanc-TFD}
\end{figure}

\begin{figure}[H]
    \centering
    \includegraphics[width=0.5\linewidth]{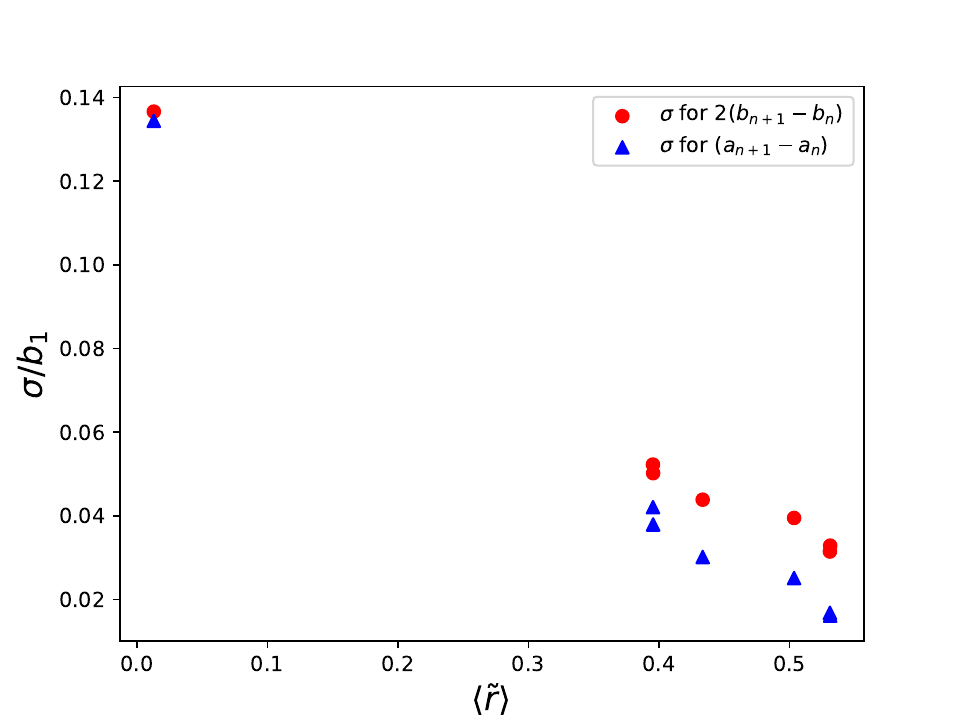}
    \caption{For initial TFD state, standard deviation of \((a_{n+1}-a_n)\) and \(2(b_{n+1}-b_n)\) plotted against average restricted ratio  \(\langle \Tilde{r} \rangle\). Such anti-correlation was observed in \cite{Balasubramanian:2024ghv}.}
    \label{fig:L-Var1}
\end{figure}

\begin{figure}[H]
    \centering
    \includegraphics[width=0.3\linewidth]{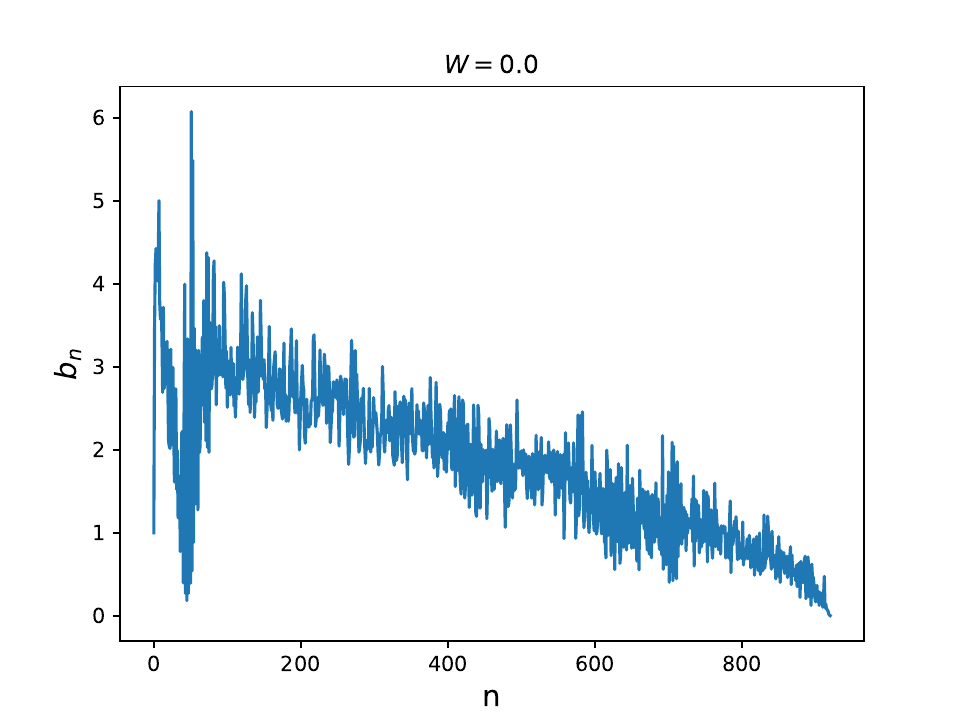}
     \includegraphics[width=0.3\linewidth]{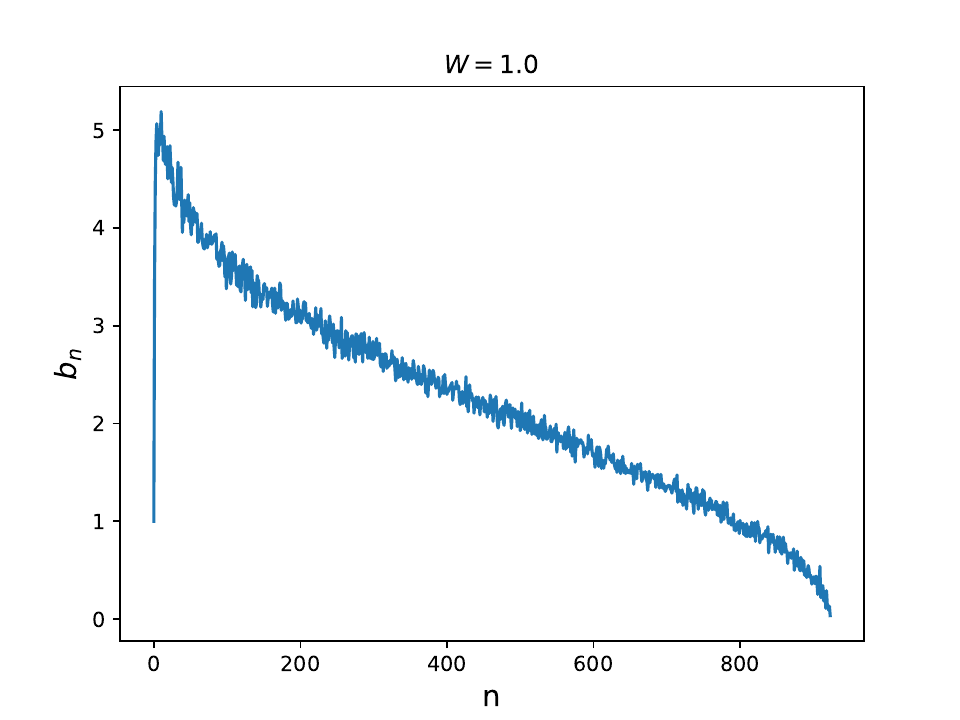}
      \includegraphics[width=0.3\linewidth]{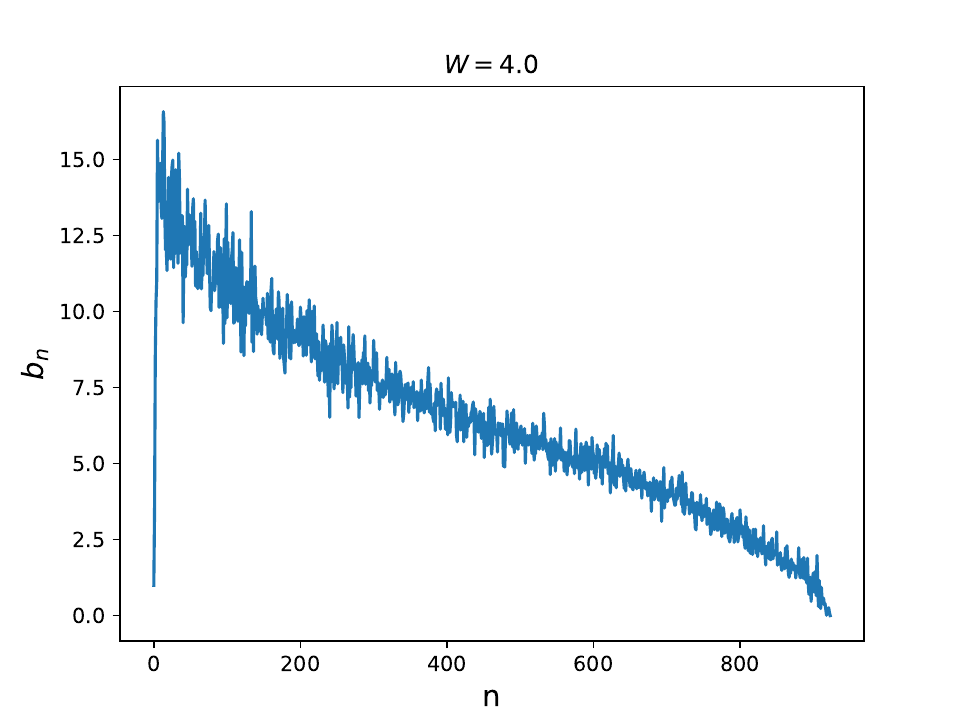}
      \vskip\baselineskip
      \includegraphics[width=0.3\linewidth]{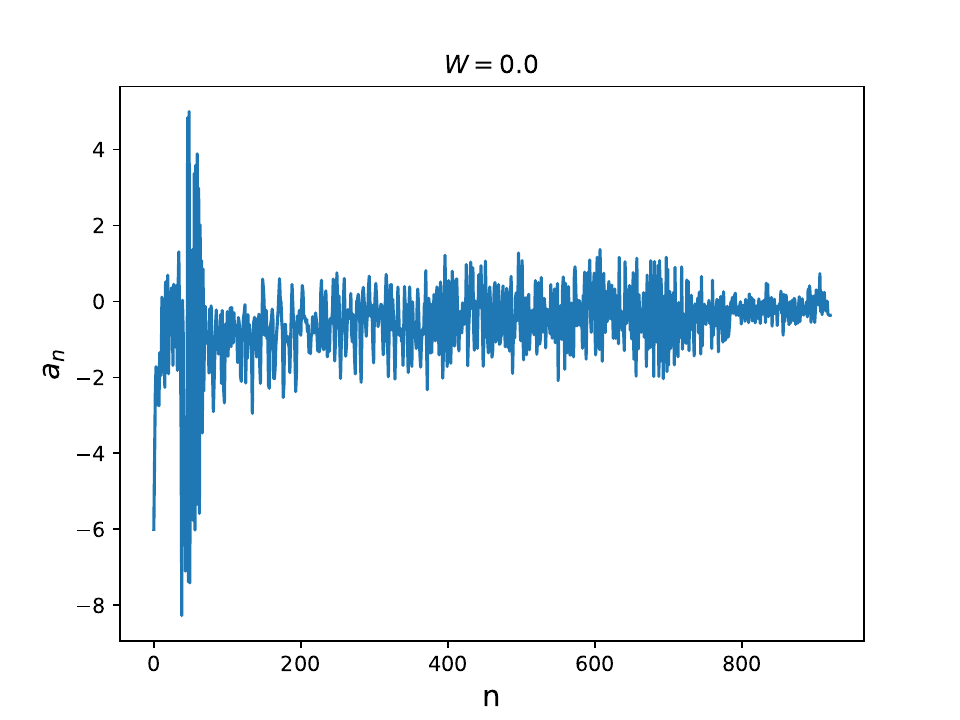}
     \includegraphics[width=0.3\linewidth]{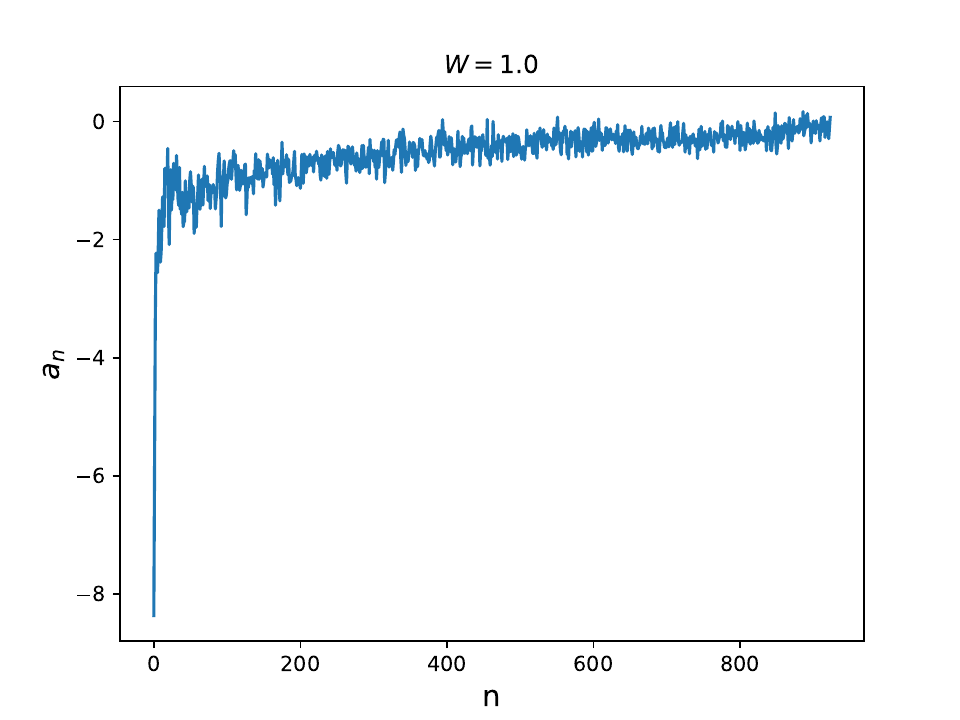}
      \includegraphics[width=0.3\linewidth]{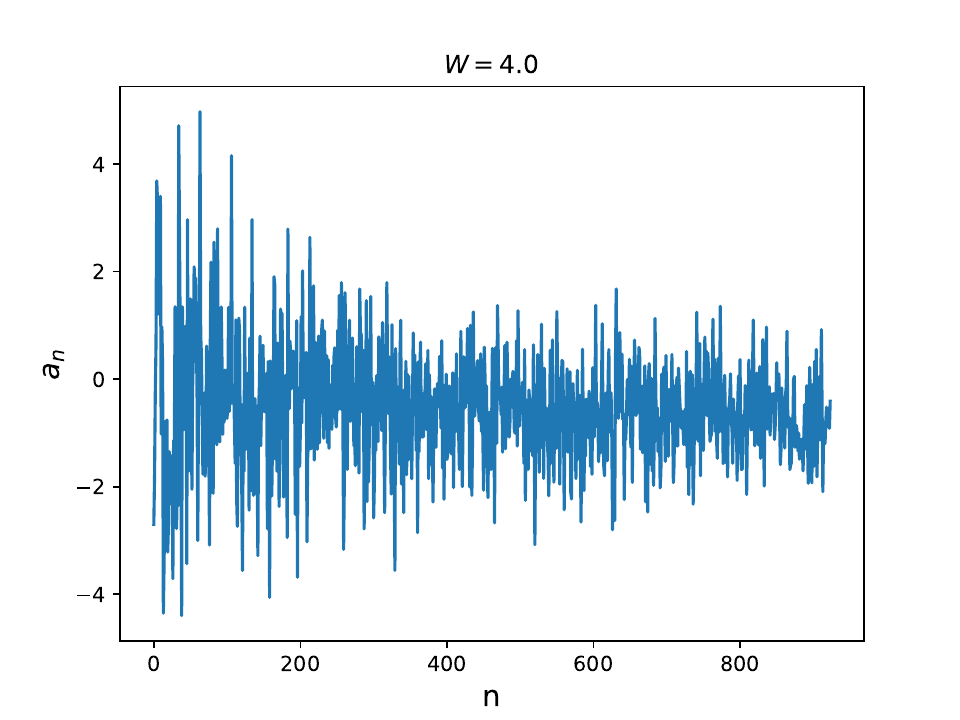}
    \caption{Lanczos coefficients with initial N\'eel state. Top-panel: \(b_n\) vs. \(n\) for different disorder strengths. Bottom-panel: \(a_n\) vs. \(n\) for different disorder strengths. \(L=12\) }
    \label{fig:Lanc-Neel}
\end{figure}

\begin{figure}[H]
    \centering
    \includegraphics[width=0.5\linewidth]{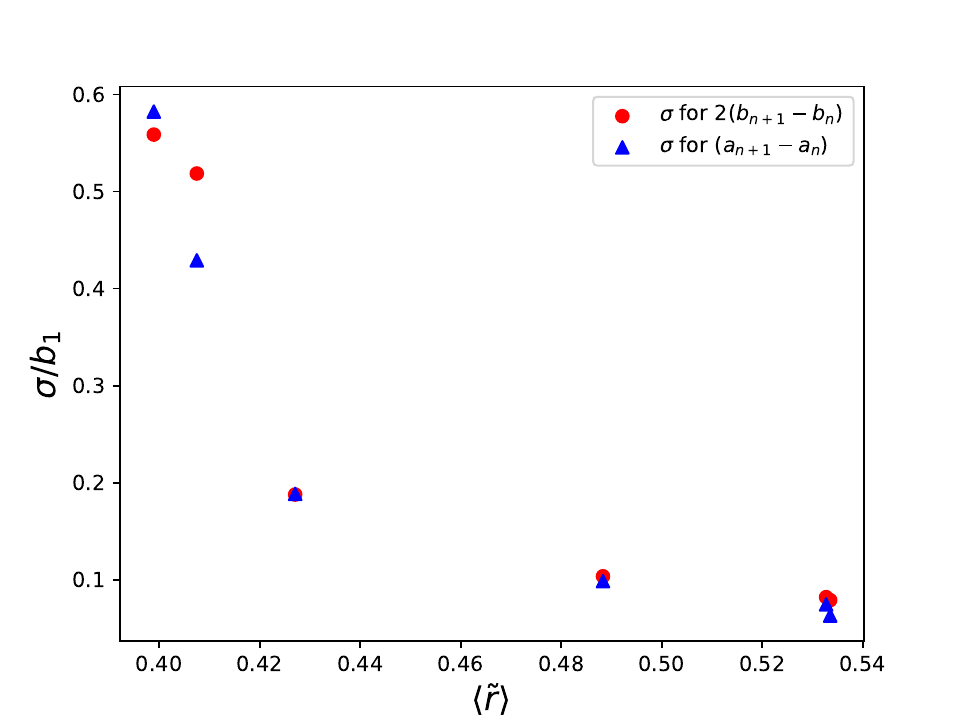}
    \caption{For initial N\'eel state, standard deviation of \((a_{n+1}-a_n)\) and \(2(b_{n+1}-b_n)\) plotted against average restricted ratio  \(\langle \Tilde{r} \rangle\). }
    \label{fig:L-Var2}
\end{figure}

The variance of Lanczos coefficients also captures chaotic and integrable behavior. In particular, integrable systems will have a higher variance of Lanczos coefficients, and chaotic systems will have comparatively less variance of Lanczos coefficients, see Fig. \ref{fig:Lanc-TFD}, Fig. \ref{fig:Lanc-Neel}. This is observed in the case of operator growth also. In the context of the spread complexity of TFD state in random matrix theory (RMT) ensembles, the variance of Lanczos coefficients has anti-correlation with the average level-spacing ratio \cite{Balasubramanian:2024ghv}. This holds for our model also; see Fig. \ref{fig:L-Var1} for the initial TFD state and Fig. \ref{fig:L-Var2} for the initial N\'eel state.

\section{Eigenstate IPR and Spectral Form Factor}

\begin{figure}[]
    \centering
    \includegraphics[width=0.35\linewidth]{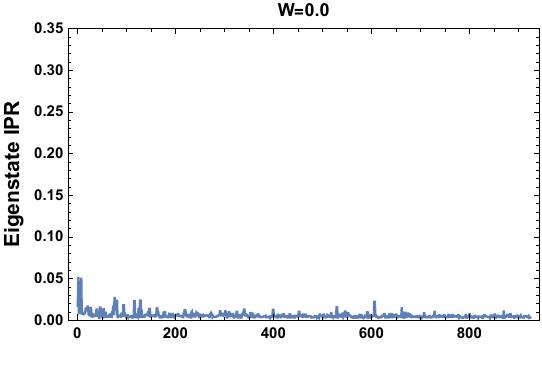}
    \includegraphics[width=0.35\linewidth]{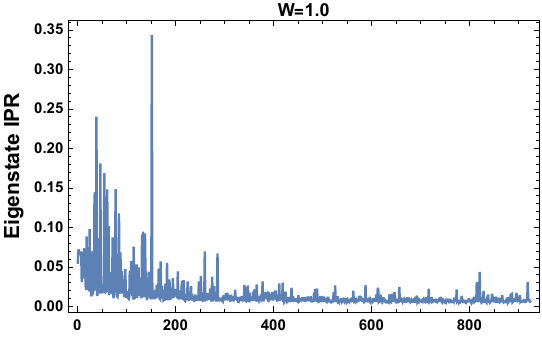}
    \vskip\baselineskip
    \includegraphics[width=0.35\linewidth]{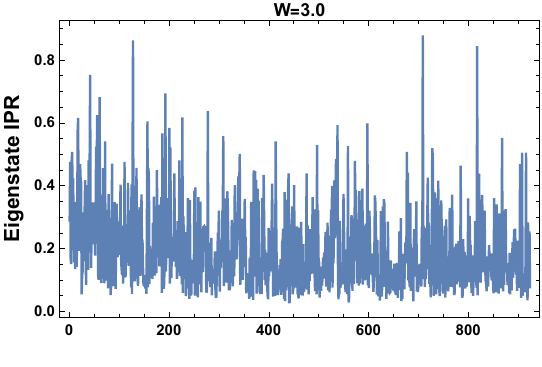}
    \includegraphics[width=0.35\linewidth]{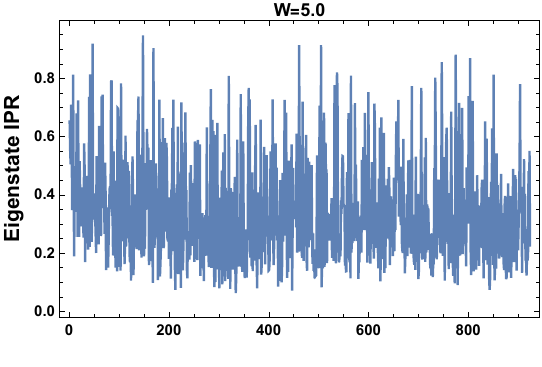}
    
    \caption{Eigenstate IPR for different values of disorder strength. With increasing \(W\), the eigenstates become more and more localized.}
    \label{fig:eig IPR}
\end{figure}

One of the important characteristics of the many-body localized phase is the localization of many-body eigenstates in real space due to the presence of disorder. This can be quantified by evaluating the inverse participation ratio (IPR) of many-body eigenstates \(\ket{\Psi_n}\) in the computational basis elements \(\ket{i}\), \(\text{IPR} = \sum_i |\bra{\Psi_n}i\rangle|^4\). A higher value of IPR will capture the localization of the eigenstates. In Fig. \ref{fig:eig IPR} we have plotted the IPR of many-body eigenstates for different disorder strengths. It is observed that the disorder-free integrable phase has the most delocalized eigenstates; on the contrary, the MBL phase has the most localized eigenstates. Therefore, we observe that the localization of eigenstates increases monotonically with the disorder strength. 

While the eigenstate IPR sheds light on the localization property of the MBL phase, it does not capture the emergent integrability aspect. Emergent integrability can be understood from the level-spacing ratio as well as spectral form factor (SFF). Spectral form factor is defined by,
\begin{equation}
\text{SFF}(t) = \frac{1}{d^2}\sum_{n,m=1}^de^{i(E_n-E_m)t}
\end{equation}

\begin{figure}[]
    \centering
    \includegraphics[width=0.3\linewidth]{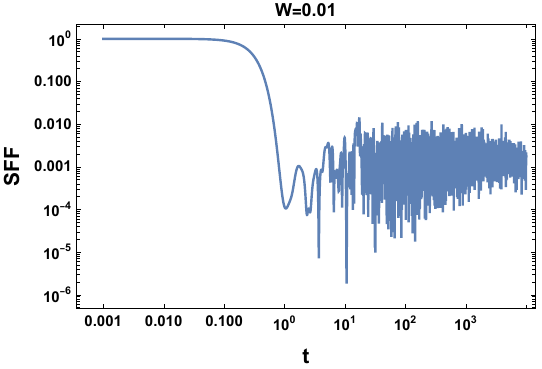}
    \includegraphics[width=0.3\linewidth]{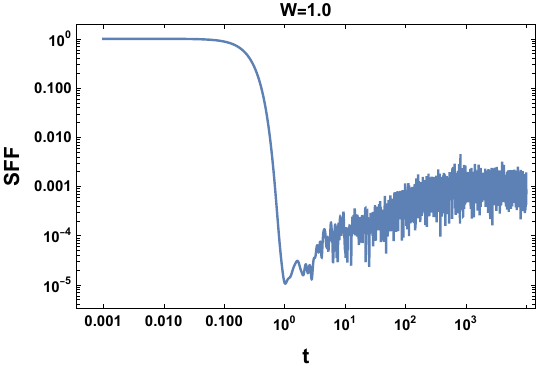}
    \vskip\baselineskip
    \includegraphics[width=0.3\linewidth]{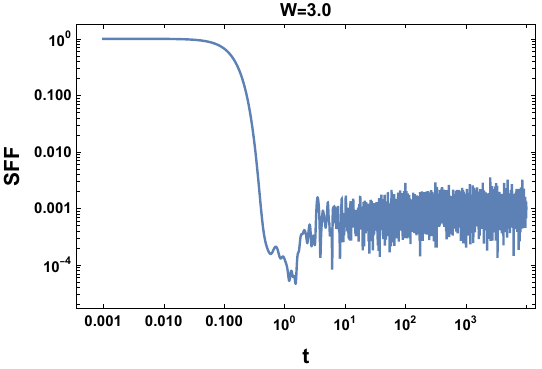}
    \includegraphics[width=0.3\linewidth]{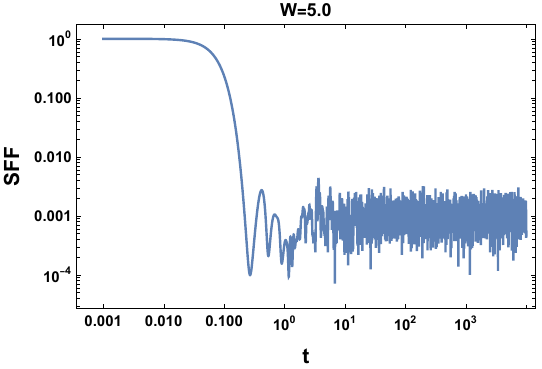}
    
    \caption{Spectral form factor (SFF) for different values of disorder strength. In the ergodic phase, we can observe a characteristic dip-ramp-plateau behavior, which is absent in low disorder and in the MBL phase.}
    \label{fig: SFF}
\end{figure}

For chaotic eigenvalue distribution, the SFF shows a characteristic dip-ramp-plateau behavior as a function of time \(t\). However, this feature is absent in integrable models. From Fig. \ref{fig: SFF}, we observe that the ergodic phase SFF has indeed a dip-ramp-plateau behavior, which is absent in both the disorder-free integrable phase and the MBL phase.

\section{Quench Dynamics}
\begin{figure}[]
    \centering
    \includegraphics[width=0.6\linewidth]{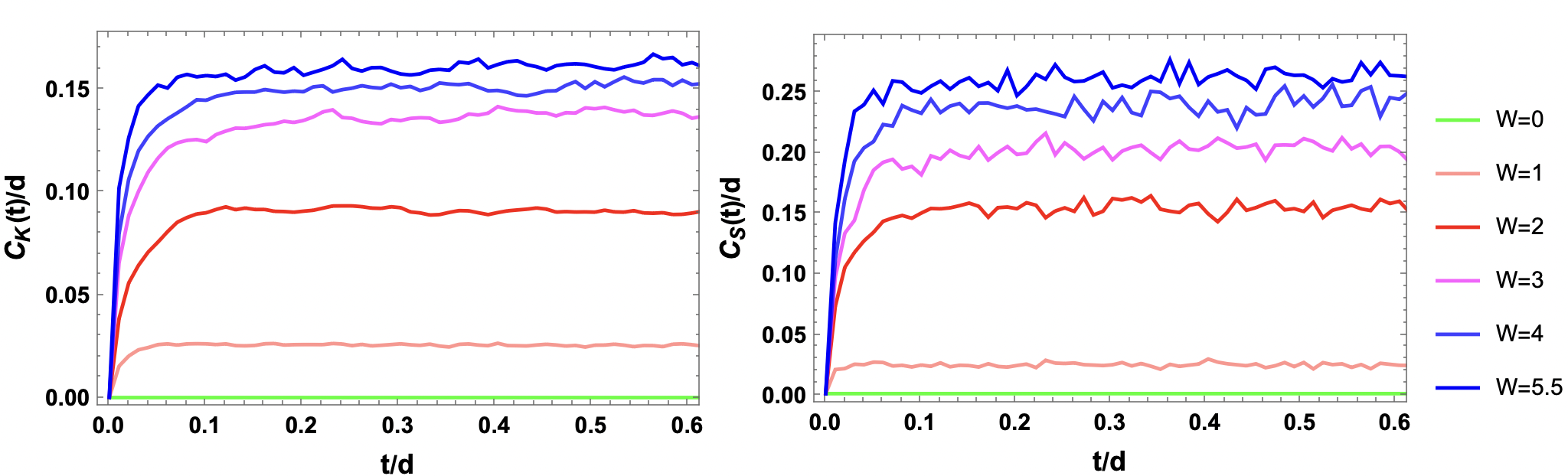}
    \vskip\baselineskip
    \includegraphics[width=0.6\linewidth]{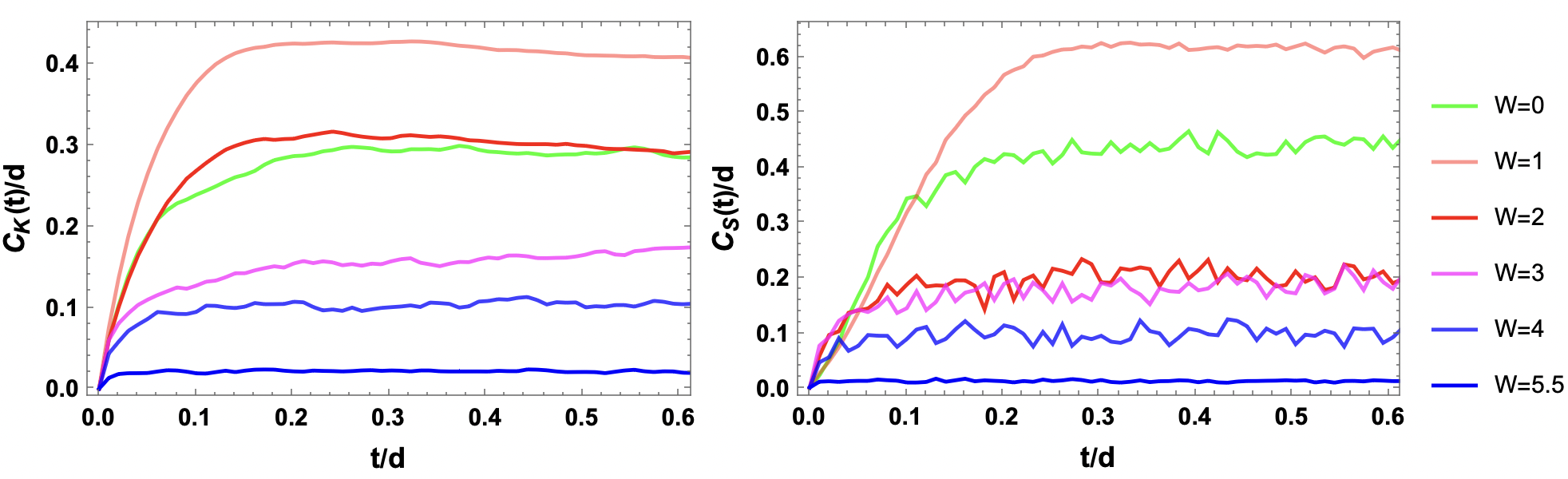}
    \vskip\baselineskip
    \includegraphics[width=0.6\linewidth]{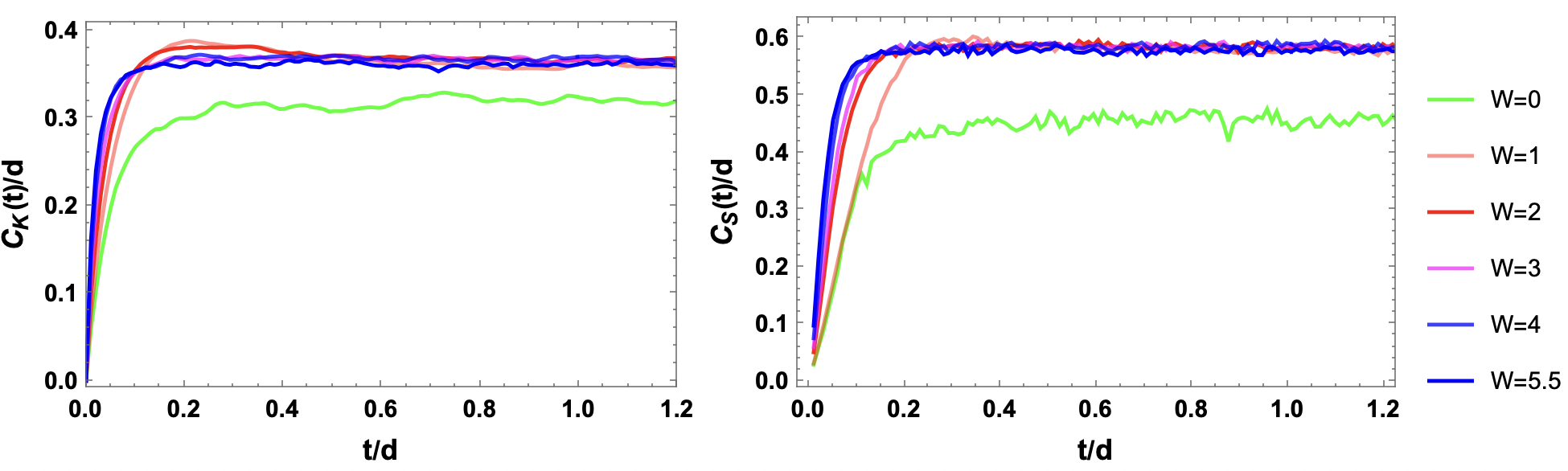}
    \caption{\textbf{Top-panel}: quenching the ground state of \(W=0\) by increasing disorder. \textbf{Middle-panel}: quenching the GS of \(W=6.5\) by decreasing disorder. Bottom-panel: quenching the TFD (\(\beta=0\)) state of \(W=0\) by increasing disorder. We have used the result for system size \(L=12\), where \(d=924\).}
    \label{fig:quench-rev.quench-quench.TFD}
\end{figure}

we consider some possible quench scenarios to understand state-dependent spread complexity in different phases. To do so at first we assume the disorder-free phase to be at its ground state, then suddenly we add disorder in the system with varying strengths. Moreover, we track the complexity dynamics of the state under the evolution of the disordered Hamiltonian. Our observation indicates that the complexity and delocalization both increase with increasing strength of disorder, and the normalized spread complexity seems to saturate to around \(0.16\) towards high disorder (top-panel in Fig. \ref{fig:quench-rev.quench-quench.TFD}). In contrast, if one starts with an infinite temperature TFD state corresponding to the disorder-free model and evolves with disordered Hamiltonian, then in the high disorder limit, the normalized complexity seems to saturate to around \(0.36\) (bottom-panel in Fig. \ref{fig:quench-rev.quench-quench.TFD}).

Further, to understand the reversed quench scenario, we take the initial state, which is the ground state of a phase inside the MBL regime, say \(W=6.5\), and evolve it with a Hamiltonian with lesser disorder strength. At \(W=6.5\), it will have zero complexity since the complexity evolution of eigenstates is trivial. However, for lesser disorders, the state will evolve non-trivially, and we observe that the complexity saturation increases with the decrement of $W$. In our case for system size $L = 12$ as we go away from \(W=6.5\), complexity increases with a maximum saturation around \(0.4\), but as we enter the disorder-free phase, the complexity saturation decreases to around \(0.29\) for \(W=0\) (middle-panel in Fig. \ref{fig:quench-rev.quench-quench.TFD}).

Taking into account the observations and previous results on thermofield dynamics, relaxation dynamics and quench dynamics, it gives an indication that the disorder-free phase generally exhibits lower complexity compared to the other two phases. The results for typical state spread complexity in the main text also support this observation.

\end{document}